\documentclass[a4paper,fleqn,usenatbib]{mnras}

\usepackage[T1]{fontenc}
\usepackage{ae,aecompl}

\usepackage{graphicx}	
\usepackage{amsmath}	
\usepackage{amssymb}	
\usepackage{multirow}
\usepackage{booktabs}
\usepackage{rotating}
\usepackage{xcolor}

\newcommand{\integral}{\textit{INTEGRAL}}

\newcommand{\nicer}{\textit{NICER}}
\newcommand{\nustar}{\textit{NuSTAR}}

\newcommand{\swift}{{\it Swift}}

\newcommand{\maxi}{MAXI~J1820+070}

\title[\maxi\ in the hard state with \nustar]{\maxi\ with \nustar\ I. An increase in variability frequency but a stable reflection spectrum: coronal properties and implications for the inner disc in black hole binaries}

\author[D. J. K. Buisson et al.]{D. J. K. Buisson$^{1}$,\thanks{Email: djkb2@ast.cam.ac.uk}
	A. C. Fabian$^{1}$,
	D. Barret$^{2,3}$,
	F. F{\"u}rst$^{4}$,
	P. Gandhi$^{5}$,
\newauthor	J. A. Garc\'ia$^{6,7}$,
	E. Kara$^{8}$,
    K. K. Madsen$^{6}$,
    J. M. Miller$^{9}$,
	M. L. Parker$^{4}$,
\newauthor	A. W. Shaw$^{10}$,
    J. A. Tomsick$^{11}$
    and D. J. Walton$^{1}$\\
  $^{1}$Institute of Astronomy, Madingley Road, Cambridge, CB3 0HA\\
  $^{2}$Universit\'e de Toulouse; UPS-OMP; IRAP; Toulouse, France\\
  $^{3}$CNRS; Institut de Recherche en Astrophysique et Plantologie; 9 Av. colonel Roche, BP 44346, F-31028 Toulouse cedex 4, France\\
  $^{4}$European Space Astronomy Centre (ESA/ESAC), E-28691 Villanueva de la Ca\~nada, Madrid, Spain\\
  $^{5}$Department of Physics and Astronomy, University of Southampton, Highfield, Southampton, SO17 1BJ\\
  $^{6}$Cahill Center for Astrophysics, 1216 E. California Blvd, California Institute of Technology, Pasadena, CA 91125, USA\\
  $^{7}$Dr. Karl Remeis-Observatory and Erlangen Centre for Astroparticle Physics, Sternwartstr. 7, 96049 Bamberg, Germany\\
  $^{8}$Department of Astronomy, University of Maryland, College Park, MD 20742-2421, USA\\
  $^{9}$Department of Astronomy, University of Michigan, 1085 South University Avenue, Ann Arbor, MI 48109-1104, USA\\
  $^{10}$Department of Physics, University of Alberta, CCIS 4-181, Edmonton, AB T6G 2E1, Canada\\
  $^{11}$Space Sciences Laboratory, 7 Gauss Way, University of California, Berkeley, CA 94720-7450, USA\\
}

\date{In review. Received 2019 September 4; in original form 2019 May 31}

\pubyear{2019}

\begin{document}
\label{firstpage}
\pagerange{\pageref{firstpage}--\pageref{lastpage}}
\maketitle

\begin{abstract}
\maxi\ (optical counterpart ASASSN-18ey) is a black hole candidate discovered through its recent very bright outburst.
The low extinction column and long duration at high flux allow detailed measurements of the accretion process to be made.
In this work, we compare the evolution of X-ray spectral and timing properties through the initial hard state of the outburst.
We show that the inner accretion disc, as measured by relativistic reflection, remains steady throughout this period of the outburst.
Nevertheless, subtle spectral variability is observed, which is well explained by a change in coronal geometry.
However, characteristic features of the temporal variability -- low-frequency roll-over and QPO frequency -- increase drastically in frequency, as the outburst proceeds.
This suggests that the variability timescales are governed by coronal conditions rather than solely by the inner disc radius.
We also find a strong correlation between X-ray luminosity and coronal temperature. This can be explained by electron pair production with a changing effective radius and a non-thermal electron fraction of $\sim20$\%.
\end{abstract}

\section{Introduction}

The transfer of matter in accretion produces variability on all timescales from the complete transfer of matter down to the shortest associated with the system.
In black hole binaries (BHBs), accretion occurs onto a particularly compact object, so timescales are correspondingly short and accretion episodes can evolve quickly (compared to, for example, active galactic nuclei, AGN). BHBs are therefore ideal laboratories for observations of long-timescale accretion processes.

X-ray emission from BHBs occurs principally in two accretion states (along with some additional transitional states), commonly referred to as soft and hard \citep[e.g. review by][]{remillard06}.
In the soft state, emission is dominated by pseudo-blackbody thermal emission from the disc \citep{novikov73,shakura73}, which extends to the innermost stable circular orbit \citep[ISCO,][]{gierlinski04,steiner10}.
Contrastingly, hard state emission is dominated by coronal emission produced by inverse-Compton scattering in a cloud of hot electrons \citep{thorne75,sunyaev79}, which has a spectrum approximated by a powerlaw with a high-energy cut-off.

A complete understanding of the physical changes between these two states is not yet well known: in particular, the nature of the inner disc during the hard state is still not agreed upon.
In some models, the disc is truncated and the accreting material forms a hot inner flow with high ionisation, which produces the Comptonised spectrum \citep{esin97,done07,gilfanov10}.
However, this is sometimes at odds with the inner radius measured spectrally, which is often small \citep{park04,reis13,parker15}.
In this case, the central part of the disc is cool and dense enough to reflect but only emits a small fraction of the energy released by accretion thermally \citep{reis10}, as energy is extracted magnetically to power a corona positioned above the disc, possibly as the base of a jet \citep[e.g.][]{markoff05,fabian12}.

X-ray emission from BHBs also shows fast variability on many timescales. Often, specific frequencies show stronger variability, known as Quasi Periodic Oscillations (QPOs; e.g. \citealt{vanderklis06}). These QPOs give characteristic timescales to the system's variability, so can be used to infer physical properties when combined with theoretical models for their production.

QPOs can be divided into various classes; the primary distinction being between high-frequency (HF, $\sim10-10^3$\,Hz) and low-frequency (LF, $\sim10^{-2}-10$\,Hz) QPOs. Low-frequency QPOs are further divided into subtypes depending on their coherence and the strength of different harmonics \citep{wijnands99,homan01,remillard02}.
Unfortunately, there is not yet an accepted explanation for the production of any of the classes of QPOs. 
In the hot inner flow model, the boundary between the disc and the hot inner flow provides a possible source of QPOs. The inner flow can undergo Lense-Thirring precession, with frequencies similar to those seen in low-frequency QPOs \citep{stella99,ingram09,ingram11}.

New observations of bright sources with the new generation of telescopes have the potential to resolve these questions.

\subsection{\maxi}

\maxi\ is a recently discovered transient source, which is likely to be a black hole binary system.
The optical counterpart to \maxi, ASASSN-18ey, was detected by the All-Sky Automated Search for SuperNovae \citep{shappee14} on 2018 March 3, several days before the announcement of the X-ray source \citep{kawamuro18} and their association was proposed \citep{denisenko18} on 2018 March 11.
The low extinction column and long outburst have allowed a wealth of data to be collected in many wavebands.

\begin{figure}
\centering
\includegraphics[ trim={0.cm 0 0 0},width=\linewidth]{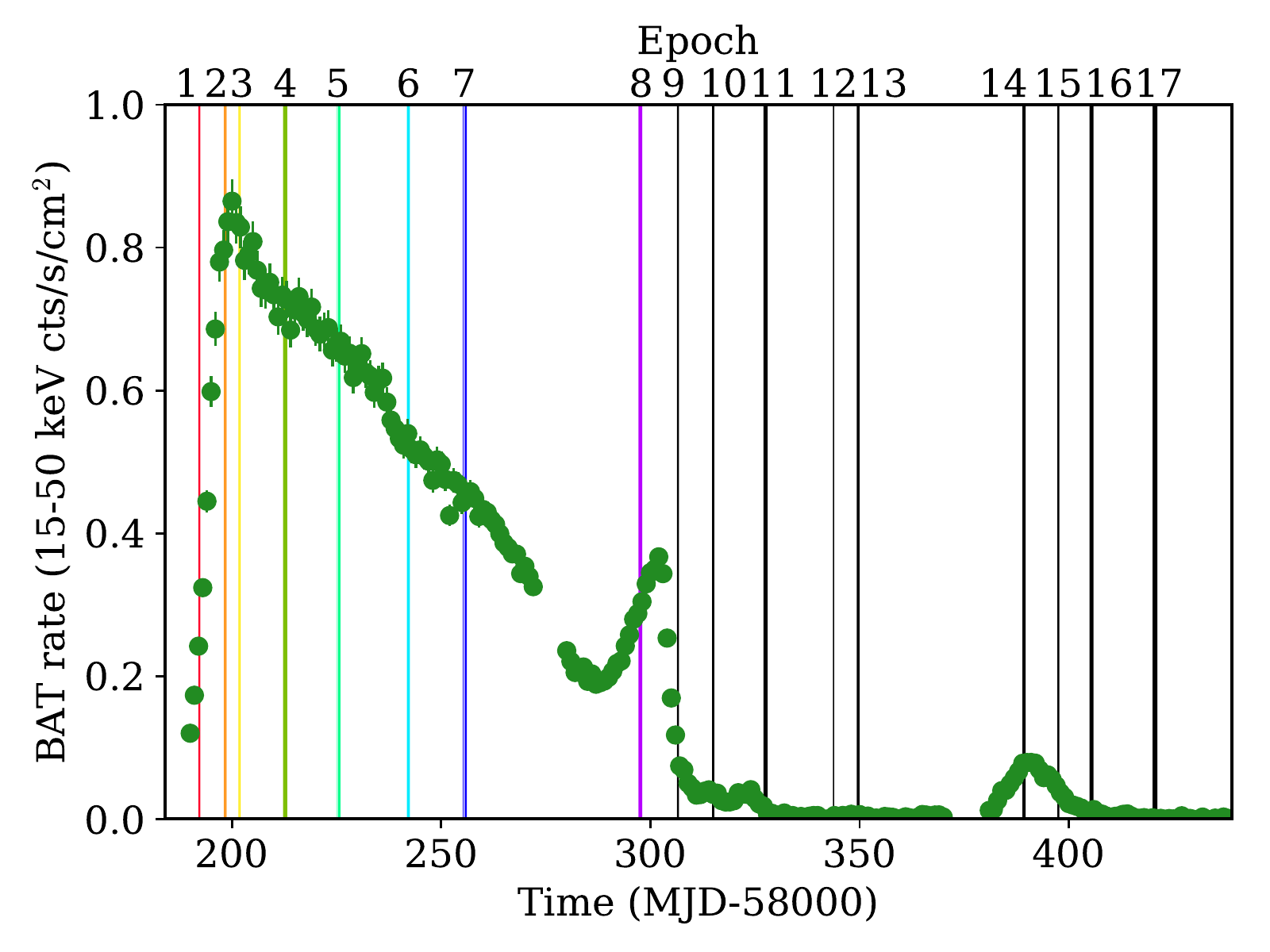}
\caption{Light curve of \maxi\ from \swift-BAT (green) with times of \nustar\ observations shown as vertical bars. Observations analysed here are in colour; later observations are in black
}
\label{fig_lc_comp}
\end{figure}

\begin{figure}
\centering
\includegraphics[ trim={0.cm 0 0 0},width=\linewidth]{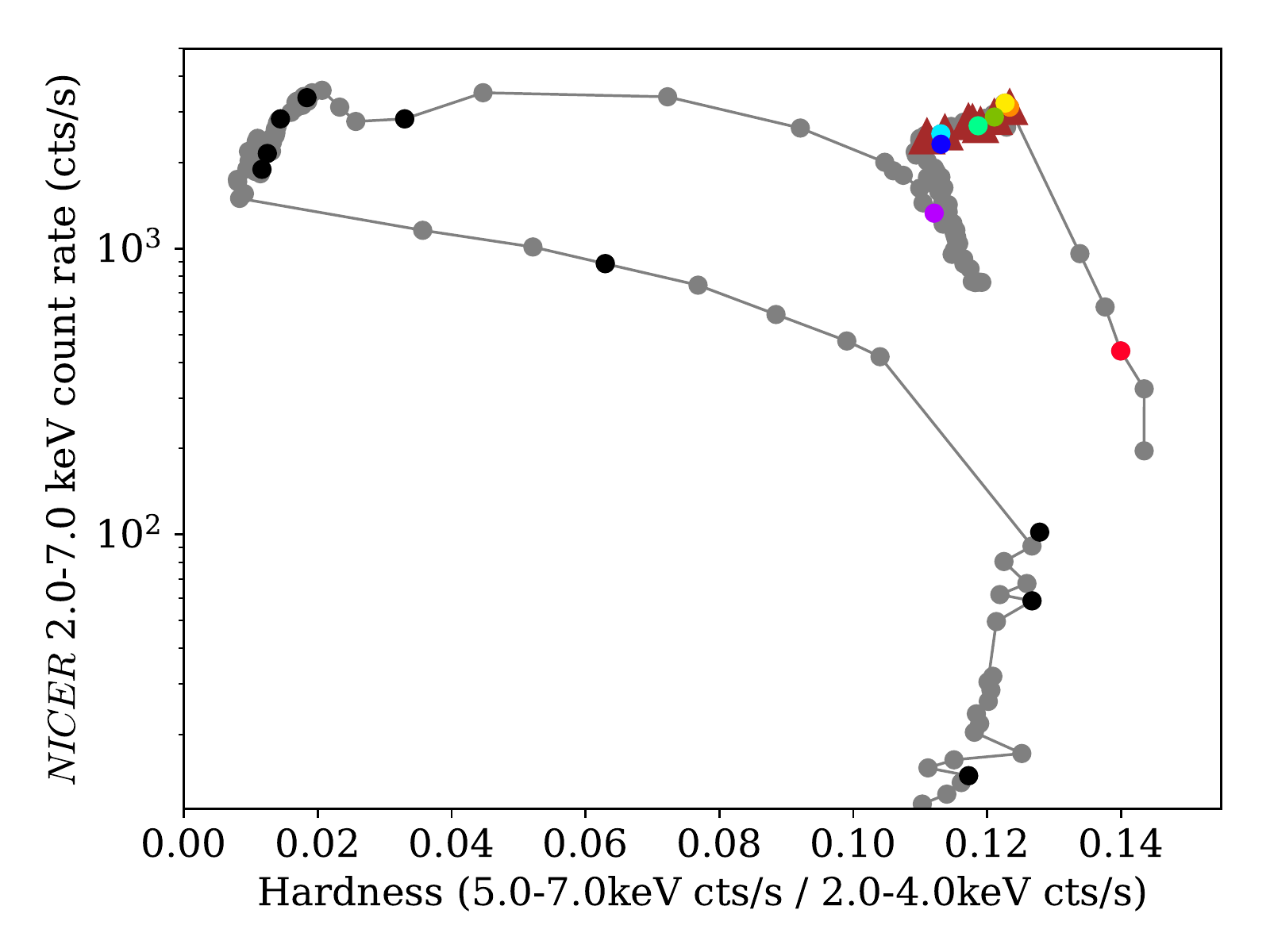}
\caption{
Hardness-intensity diagram of \maxi\ from \nicer\ data (grey). The day coincident with each \nustar\ observation is shown in the colour matching Figure~\ref{fig_lc_comp}. For comparison, the \nicer\ observations analysed in \citet{kara19} are shown as brown triangles.
}
\label{fig_hid}
\end{figure}

The X-ray outburst began with an initial fast rise (to MJD 58200) and slow decay (till around MJD 58290) across the full X-ray band; once the flux reached roughly one quarter of the peak, the source re-brightened substantially before the hard X-ray flux dropped dramatically (around MJD 58305) as the source transitioned into the soft state. After over 2 months in the soft state, the hard X-ray flux increased (from around MJD 58380) as the source re-entered the hard state before fading into quiescence. Figure~\ref{fig_lc_comp} shows the hard ($15-50$\,keV) X-ray light curve from the \textit{Neil Gehrels Swift Observatory} Burst Alert Telescope (\swift-BAT) transient monitor \citep{gehrels04,krimm13}. Figure~\ref{fig_hid} shows a hardness-intensity diagram of the outburst from \textit{Neutron star Interior Composition ExploreR} \citep[\nicer;][]{gendreau16} data.

\textit{International Gamma Ray Astrophysics Laboratory} \citep[\integral;][]{winkler03} observations show differences in the spectra between the rise and decay and a hard tail above the Compton cut-off, which may be from a jet \citep{roques19}.
Combining \textit{Monitor of All-sky X-ray Image} \citep[\textit{MAXI};][]{matsuoka09} with \swift-BAT data shows a typical photon index of $\Gamma\sim1.5$ and an electron temperature of $kT_{\rm e}\sim50$\,keV \citep{shidatsu18}.

The start of the optical outburst preceded the X-ray by around 7 days; lightcurves and spectra show broad double peaked emission lines and variability in the outburst and pre-outburst emission, typical of low mass X-ray binaries \citep{tucker18}. 

QPOs have been observed in the emission from \maxi\ in many wavebands, from optical \citep{yu18a,yu18b,zampieri18,fiori18} to hard X-ray \citep{mereminskiy18}.
The frequency of these QPOs increases with time \citep{homan18_freqevolution} during the first part of the outburst and, over the initial few \textit{Nuclear Spectroscopic Telescope Array} \citep[\nustar,][]{harrison13} hard X-ray observations, this increase was suggested to be exponential with time \citep{buisson18_maxi}.
Reverberation lags, differences in arrival time between direct coronal and reflected emission, have also been detected in the X-ray variability; these lags shorten as the variability frequencies increase, suggesting that the corona becomes more compact \citep{kara19}.

The distance to \maxi\ is still moderately uncertain. Among the first set of XRB distances to be derived directly from optical astrometry (rather than indirect photometric and spectroscopic methods), \textit{Gaia} measurements of the system in quiescence give a parallax of $0.31\pm0.11$\,milliarcsec, which corresponds to a distance of $3.5_{-1.0}^{+2.2}$\,kpc \citep{gandhi19}.
This should be improved in the next \textit{Gaia} data release, especially considering the long interval over which the source remained bright.

\nustar\ is the first X-ray telescope to focus hard ($\gtrsim10$\,keV) X-rays. It uses CdZnTe detectors with a triggered readout, allowing observations  of bright sources to be free of pile-up which degrades conventionally read CCDs. These capabilities have allowed 
\nustar\ to perform several observations of \maxi; the times of these are shown in Figure~\ref{fig_lc_comp} as coloured vertical bands (these colours are used to indicate the same epoch throughout this work), showing that \nustar\ observations occurred during all of these stages of the outburst.

\begin{table*}
\caption{List of \nustar\ observations of \maxi. The observation length is significantly longer than the effective exposure due to deadtime, orbital and other gaps. Since pairs of observations are sometimes closely spaced, we divide the datasets into several epochs for analysis purposes. Only epochs before the transition to the soft state are considered here.}
\label{tab:obs}
\begin{tabular}{lcrrcrr}
\hline
OBSID & Start time & \multirow{1}{1.5cm}{Observation length/ks} & \multirow{1}{2cm}{Count rate (incident cts/s)} & Live fraction & Epoch & State \\
\\
\hline
90401309002 & 2018-03-14T20:26:09 & $43.0$ & 157 & 0.62 & 1 & Hard\\
90401309004 & 2018-03-21T00:31:09 & $14.2$ & 664 & 0.28 & 2 & Hard\\
90401309006 & 2018-03-21T07:06:09 & $31.7$ & 679 & 0.28 & 2 & Hard\\
90401309008 & 2018-03-24T12:31:09 & $20.7$ & 701 & 0.27 & 3 & Hard\\
90401309010 & 2018-03-24T20:26:09 & $14.9$ & 703 & 0.27 & 3 & Hard\\
90401309012 & 2018-04-04T04:31:09 & $84.4$ & 624 & 0.29 & 4 & Hard\\
90401309013 & 2018-04-16T22:21:09 & $8.4$ & 602 & 0.3 & 5 & Hard\\
90401309014 & 2018-04-17T06:31:09 & $55.5$ & 609 & 0.3 & 5 & Hard\\
90401309016 & 2018-05-03T18:51:09 & $60.5$ & 512 & 0.34 & 6 & Hard\\
90401309018 & 2018-05-17T03:36:09 & $13.1$ & 407 & 0.37 & 7 & Hard\\
90401309019 & 2018-05-17T14:26:09 & $43.9$ & 440 & 0.37 & 7 & Hard\\
90401309021 & 2018-06-28T03:56:09 & $77.7$ & 265 & 0.5 & 8 & Hard\\
90401309023 & 2018-07-07T08:36:09 & $38.1$ & 461 & 0.33 & 9 & Soft\\
90401309025 & 2018-07-15T17:51:09 & $43.7$ & 321 & 0.39 & 10 & Soft\\
90401309027 & 2018-07-28T01:11:09 & $83.3$ & 237 & 0.45 & 11 & Soft\\
90401309029 & 2018-08-13T14:26:09 & $26.5$ & 158 & 0.54 & 12 & Soft\\
90401309031 & 2018-08-19T07:26:09 & $58.7$ & 131 & 0.58 & 13 & Soft\\
90401309033 & 2018-09-27T21:51:09 & $67.0$ & 108 & 0.68 & 14 & Hard\\
90401309035 & 2018-10-06T07:11:09 & $38.1$ & 46 & 0.81 & 15 & Hard\\
90401309037 & 2018-10-13T22:46:09 & $82.4$ & 12 & 0.9 & 16 & Hard\\
90401309039 & 2018-10-29T01:11:09 & $96.1$ & 3 & 0.93 & 17 & Hard\\
\hline

\end{tabular}

\end{table*}

The accumulated dataset is vast and a full analysis is beyond the scope of a single work. Here, we focus on a comparison between the evolution of the spectral and timing properties during the initial hard state of the outburst.
We summarise the data used in Section~\ref{section_datareduction}; present an overview of the outburst properties in Section~\ref{sec:spec} and describe details of power spectra in Section~\ref{sec:psds}. We comment on possible interpretations of our findings in Section~\ref{sec_comparison} and summarise in Section~\ref{section_conclusions}.

\section{Observations and Data Reduction}
\label{section_datareduction}

We analyse data from all \nustar\ observations of \maxi\ before the transition to the soft state, as shown in Table~\ref{tab:obs}. For data transfer reasons, some pseudo-continuous observation periods were divided into separate OBSIDs; we reduce these sections separately but treat them as a single observation for later analysis. We refer to different observations as epochs, numbered as in Table~\ref{tab:obs}.

The data were reduced with the \textsc{nustardas} pipeline, version 1.8.0 and \textsc{CALDB} version 20171002. When filtering for passages through the South Atlantic Anomaly, \texttt{"saamode"} was set to \texttt{"strict"} and \texttt{"tentacle"} to \texttt{"yes"}. Following the recommendations of the \nustar\ team, we used the status expression \texttt{"STATUS==b0000xxx00xxxx000"} to avoid source photons being spuriously flagged as \textsc{`test'} events due to the bright source.
The source region was a circle of 60\,arcsec radius centroided to the peak brightness.
We also extracted a background from a circle of 60\,arcsec radius from the area of the same chip with the lowest apparent source contamination.
However, this background flux is negligible and source-dominated across the whole bandpass (for the observations analysed here).
We group the FPMA data to a minimum signal to noise ratio of at least 50, which allows the use of $\chi^2$ statistics, and group FPMB to the same energy bins to facilitate straightforward comparison of detectors.

To properly account for the loss of exposure due to dead-time and ensure all other instrumental effects are properly accounted for, we produce light curves using the \textsc{nuproducts} software, which includes the \textsc{nulccorr} process. To fully account for dead-time, this requires that the light curve bin size is at least 1\,s.
When studying higher frequencies than this allows ($>0.5$\,Hz), we correct for dead-time using the \textsc{hendrics} package \citep{bachetti15sc,bachetti15,bachetti18}.

To indicate the magnitude of dead-time effects, mean incident count-rates and live fractions for each observation are also given in Table~\ref{tab:obs}.

\section{Results}
\label{sec:results}

\subsection{Spectral analysis}
\label{sec:spec}

\begin{figure}
\centering
\includegraphics[width=\columnwidth]{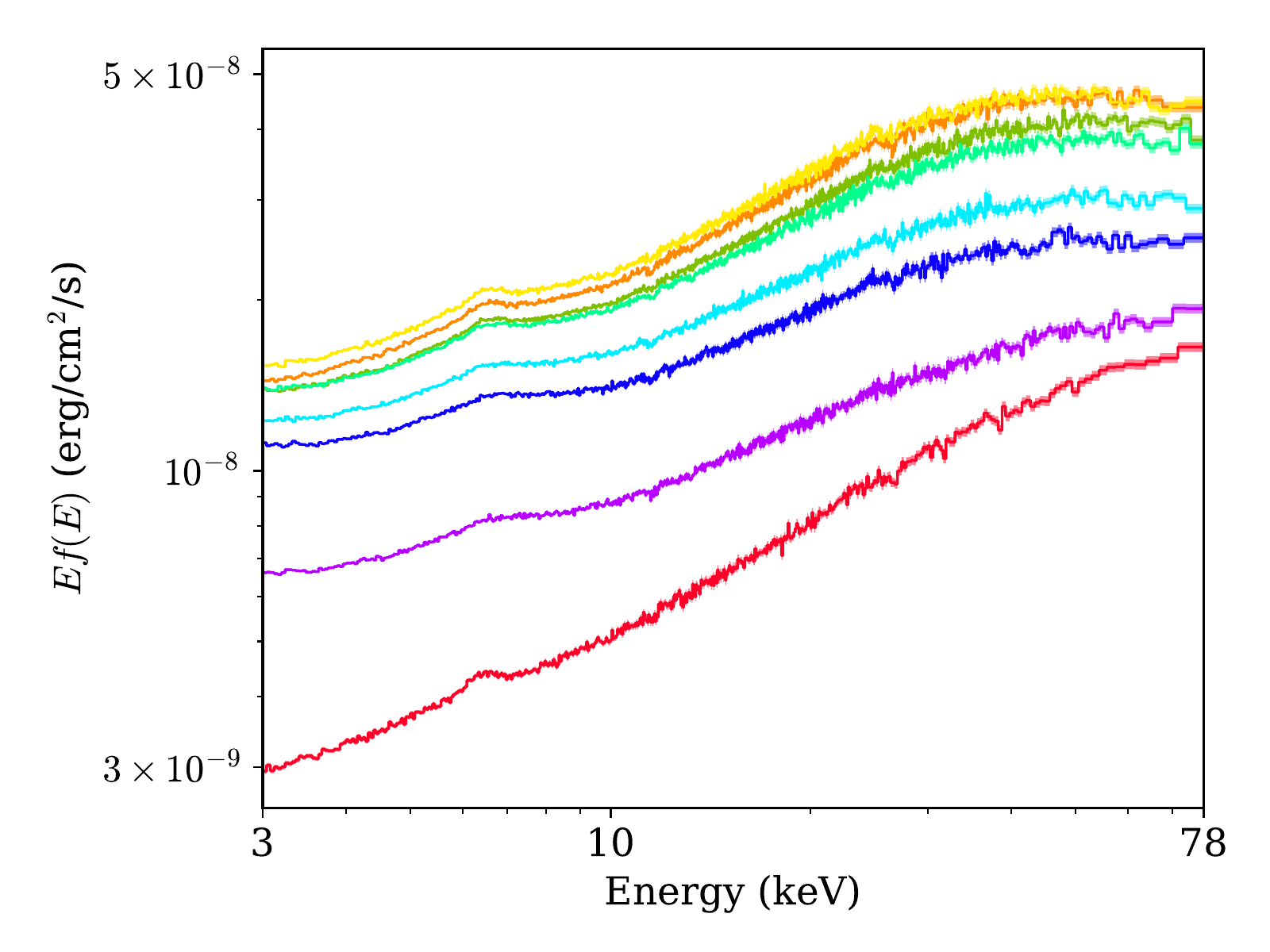}
\caption{Spectra of each \nustar\ observation unfolded to a constant model. FPMA and B have been combined for display purposes. The colour of each observation matches that in Figure~\ref{fig_lc_comp}. The source has almost constant spectral shape during the hard state, softening slowly through the initial outburst and re-hardening during the second rise.}
\label{fig_unfold}
\end{figure}

\begin{figure}
\centering
\includegraphics[width=\columnwidth]{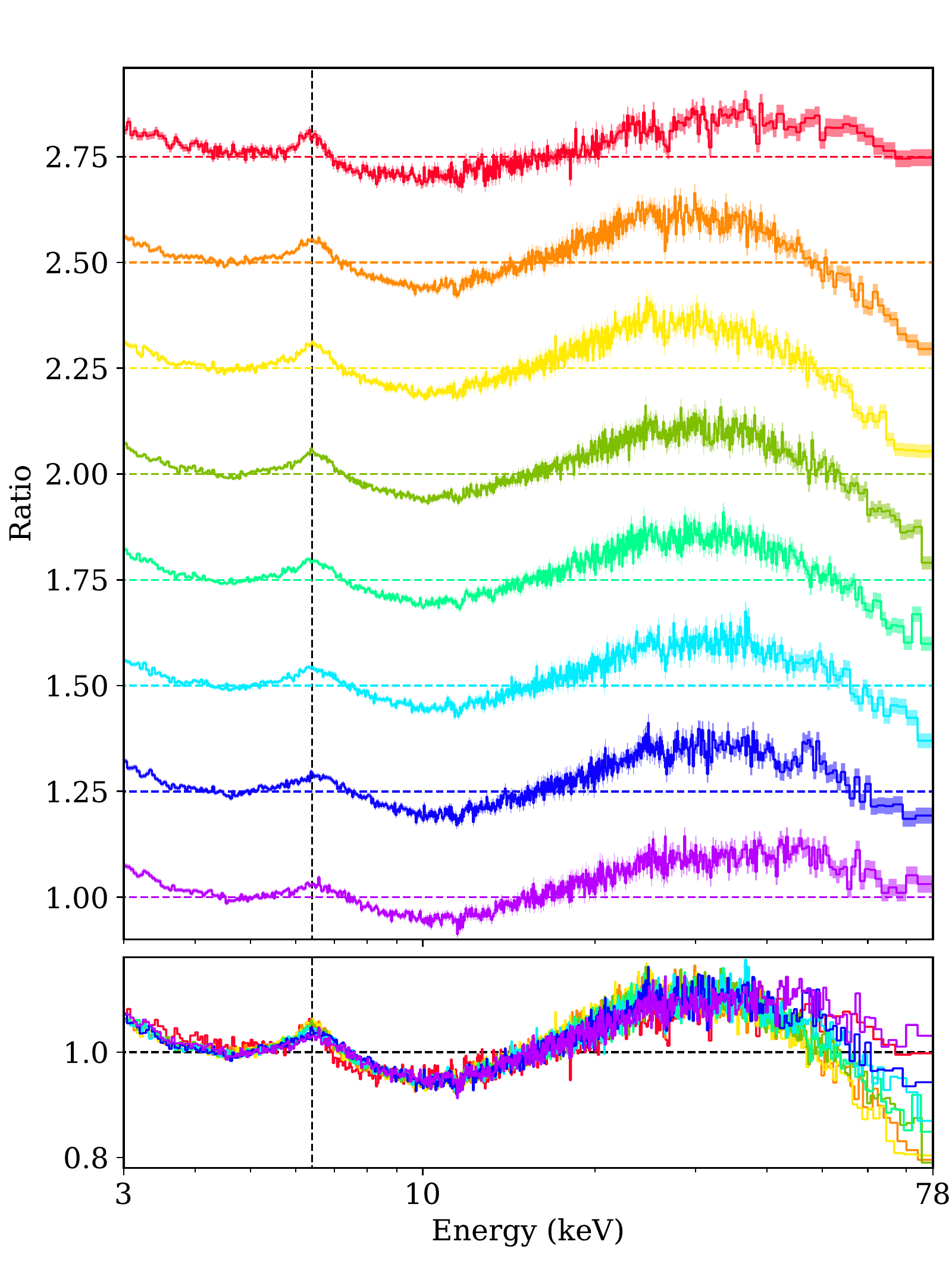}
\caption{
Ratio of the mean spectrum of each epoch in the hard state to the best-fitting powerlaw. Time runs from top to bottom in the upper panel; successive epochs are offset by 0.25, as indicated by the dashed lines. All epochs are shown superimposed in the lower panel. The colour of each observation matches that in Figure~\ref{fig_lc_comp}. The vertical dashed line indicates the rest energy ($6.4$\,keV) of the iron line. The narrow core to the iron line weakens and (apart from the first epoch) the relative high-energy flux increases throughout the outburst.
Features at $\sim12$ and $28$\,keV are calibration residuals.
}
\label{fig_nustar_ratio}
\end{figure}

\subsubsection{Qualitative comparison}

The spectrum from each epoch considered here is shown unfolded to a constant model in Figure~\ref{fig_unfold}. Apart from changes in hardness, this shows little evolution in spectral shape throughout the hard state. The spectra soften gradually till the second increase in flux (the last epoch in the hard state, epoch~8, shown in purple), when a slight hardening is seen.
To show spectral features more clearly, we also show the hard state spectra as a ratio to the best-fitting power law in Figure~\ref{fig_nustar_ratio}. This shows a broad iron K$\alpha$ emission line peaking around 6.5\,keV and a Compton hump at $20-50$\,keV, indicating the presence of relativistic reflection, as would be expected from an accretion disc extending close to a black hole \citep[e.g.][]{fabian00,reynolds03}. There is also a clear narrow core to the iron emission. The broad component of the iron line appears remarkably stable throughout the outburst, while the relative strength of the narrow core reduces with time; this behaviour is also seen in observations by \nicer\ \citep{kara19}. Additionally, the relative high energy flux increases during the outburst, possibly indicating an increase in coronal temperature.
The relative high energy flux is also significantly greater during increases in broad band flux (the first and last hard-state spectra) than decreases.

\subsubsection{Quantitative modelling}

\begin{figure}
\centering
\includegraphics[ trim={0.cm 0 0 0},width=\linewidth]{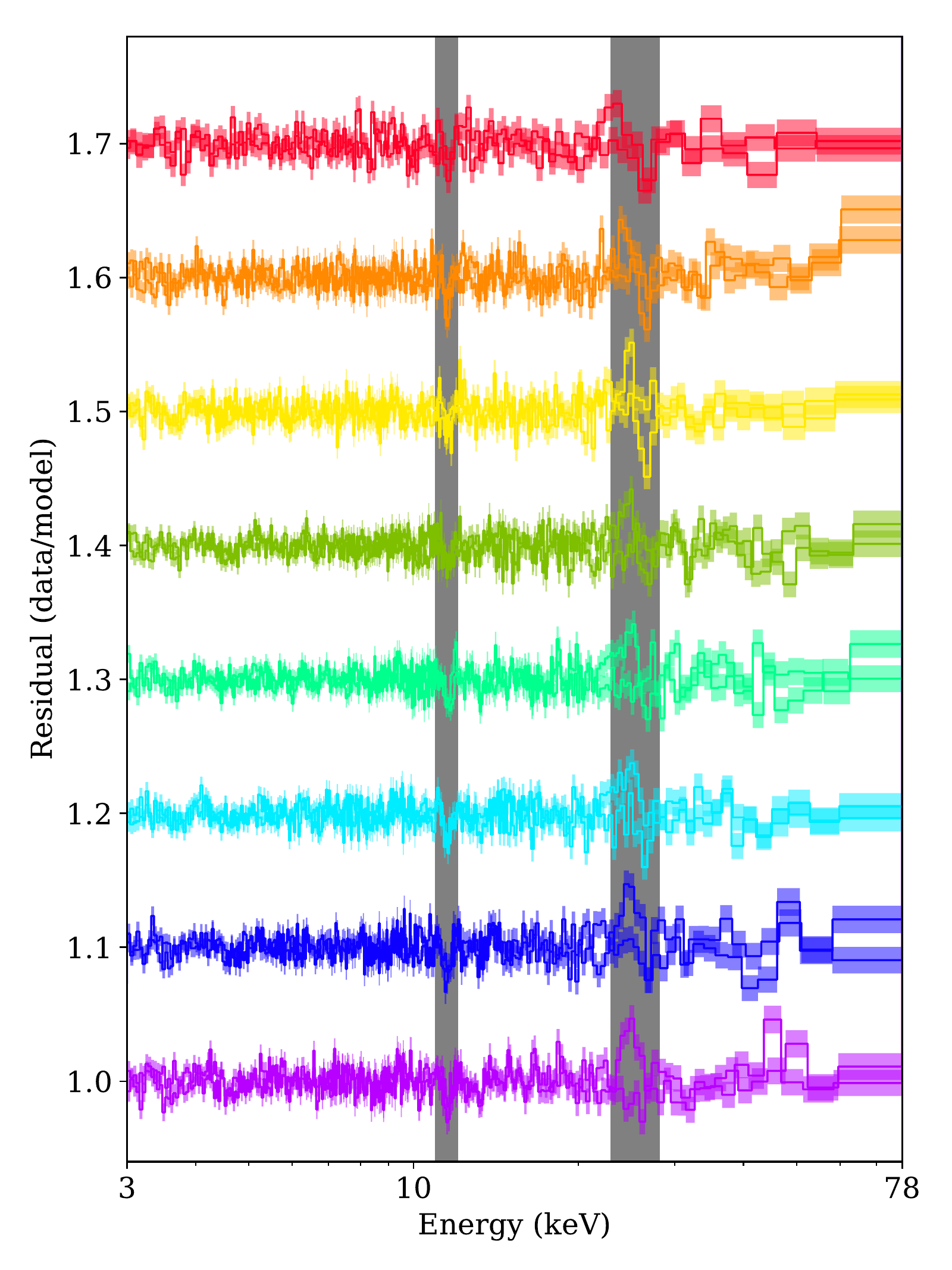}
\caption{Plot of ratio residuals to best-fit models for each spectrum. Successive spectra are offset by 0.1. The colour of each epoch matches that in Figure~\ref{fig_lc_comp}. Grey bands denote energy ranges which were ignored for fitting due to instrumental features.}
\label{fig_res}
\end{figure}

We model the hard X-ray emission as originating from a Comptonising corona illuminating a disc around a black hole. Owing to the availability of models, we make the standard geometrical approximation of a razor-thin, Keplerian disc.
From the change in iron line profile, we deduce that the illumination of the outer disc (forming the narrow core) is changing, while the illumination of the inner disc (forming the broad component) varies less. Therefore, we require an extended, changing corona. We model this simply as two point sources on the spin axis at different heights above the disc (two instances of \textsc{relxilllpCp}, \citealt{dauser10,garcia14}) with the upper point source inducing the majority of the narrow component of the reflection and the lower point source dominating the broad component. This is unlikely to be the true physical scenario (the true extension is likely continuous, especially once averaged over many dynamical times) but provides a representation with sufficient variable parameters to model the observed changes to the spectra while remaining computationally tractable.

The increase in flux at low energies relative to a simple powerlaw (see Figure~\ref{fig_nustar_ratio}) is greater than is present in the reflection in the \textsc{relxilllpCp} model (which uses \textsc{xillver}, \citealt{garcia13}). This may be due to the disc having higher density than is used in (this version of) \textsc{xillver}, which has a proton density of $n=10^{15}$\,cm$^{-3}$ as appropriate for typical AGN \citep{garcia16}. The higher density causes the reprocessed thermal continuum to move into the X-ray band \citep{garcia16,tomsick18,jiang19}. A detailed analysis of this effect requires data at softer energies than are provided by \nustar\ and will be considered in future work (Fabian et al. in prep.); here, we represent the additional soft flux with a \textsc{diskbb} component.

\begin{figure*}
\centering
\includegraphics[ trim={0.cm 0 0 0},width=\linewidth]{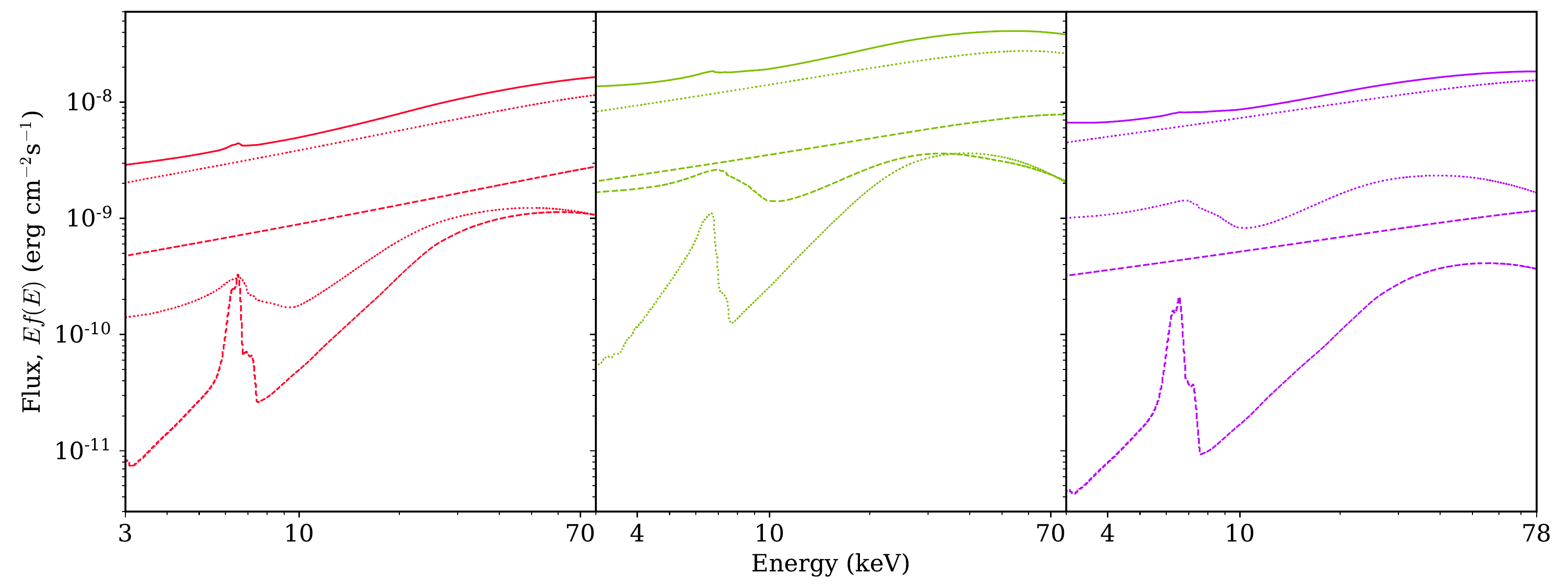}
\caption{
Plot of best-fit models to epochs 1, 4 and 8 (left to right). The colour of each epoch matches that in Figure~\ref{fig_lc_comp}. The upper line in each epoch is the total model; the contributions from the upper and lower corona are shown dashed and dotted respectively, each separated into their continuum and reflected components. The relative contribution from the narrow reflection component reduces in successive observations.
}
\label{fig_models}
\end{figure*}

\begin{figure}
\centering
\includegraphics[ trim={0.cm 0 0 0},width=\linewidth]{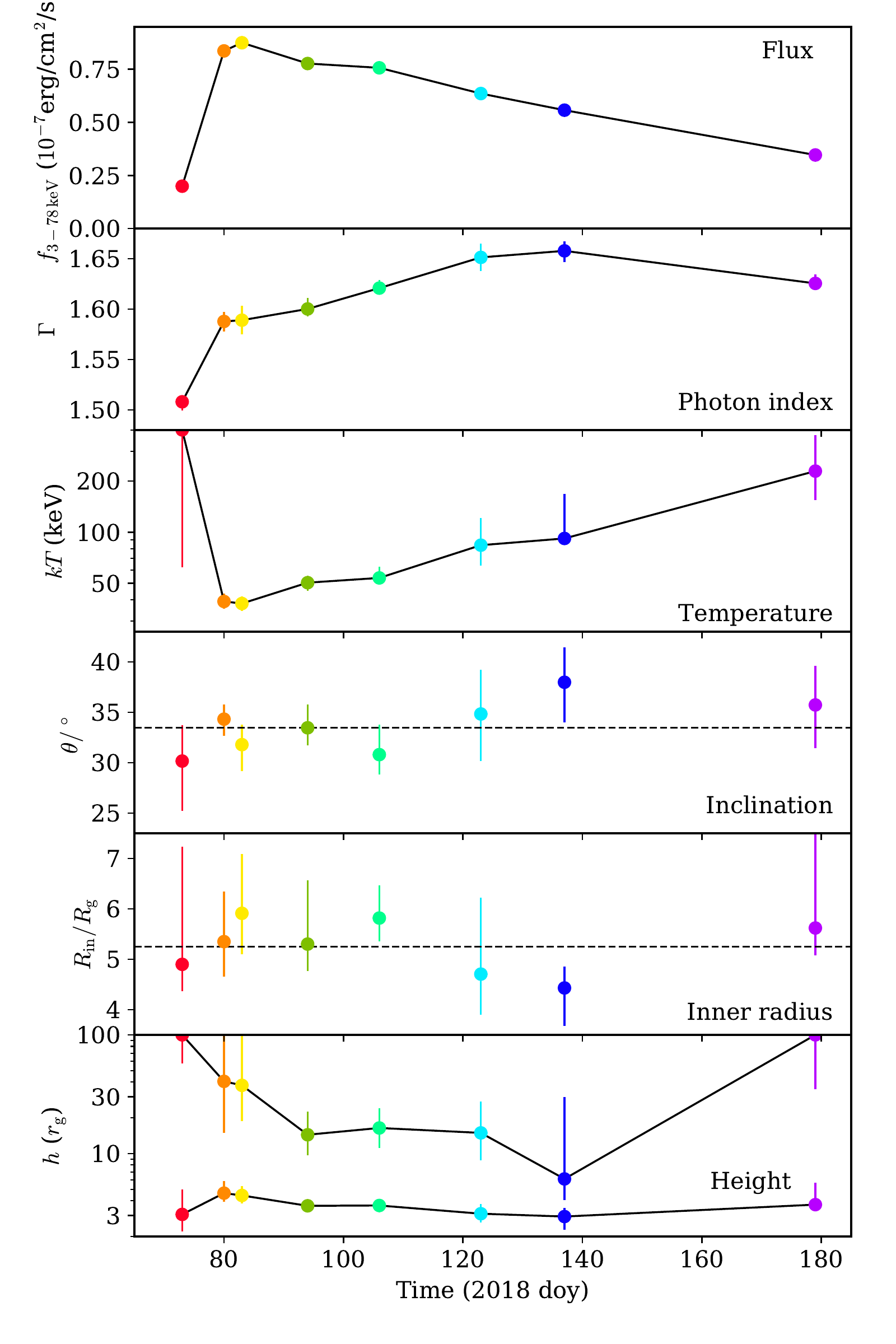}
\caption{Plot of key system parameters against time. Dashed horizontal lines indicate the weighted mean across epochs.
}
\label{fig_par_change}
\end{figure}

\begin{table*}
\caption{Parameters of fits to \maxi\ in the hard state. The model is \textsc{diskbb+relxilllpCp(1)+relxilllpCp(2)}. Errors represent 90\% confidence intervals.}
\label{tab:spec_fit}
\begin{tabular}{lllcccccccc}
\hline
Component & Model & Parameter & \multicolumn{4}{c}{Epoch} \\\cmidrule(lr){4-7}& & & 1 & 2 & 3 & 4 \\\hline\multirow{4}{1.5cm}{Soft flux} & \multirow{2}{1.5cm}{\textsc{diskbb}} & Norm$_{\rm FPMA}$ & $80_{-50}^{+40}$ & $1000_{-500}^{+800}$ & $1000_{-500}^{+700}$ & $1500_{-500}^{+900}$ \\
 &  & $kT_{\rm FPMA}$/keV & $0.8\pm0.1$& $0.68_{-0.05}^{+0.08}$ & $0.7_{-0.07}^{+0.08}$ & $0.64\pm0.05$\\
 & \multirow{2}{1.5cm}{\textsc{diskbb}} & Norm$_{\rm FPMB}$ & $300_{-300}^{+100}$ & $1500_{-900}^{+2500}$ & $2000_{-1000}^{+4000}$ & $4000_{-2000}^{+3000}$ \\
 &  & $kT_{\rm FPMB}$/keV & $0.6_{-0.2}^{+0.3}$ & $0.6_{-0.06}^{+0.09}$ & $0.58_{-0.08}^{+0.09}$ & $0.53_{-0.06}^{+0.05}$ \\
\hline\multirow{5}{1.5cm}{Compton continuum} & \multirow{5}{2.1cm}{\textsc{relxilllpCp(1/2)}} & Norm$_{\rm FPMA}$ & $0.24\pm0.07$ & $0.28_{-0.06}^{+0.08}$ & $0.32_{-0.06}^{+0.07}$ & $0.41_{-0.08}^{+0.04}$ \\
 &  & $C_{\rm FPMB/FPMA}$ & $1.023\pm0.01$& $1.03\pm0.002$& $1.031\pm0.002$& $1.029\pm0.002$\\
 &  & $\Gamma_{\rm FPMA}$ & $1.508_{-0.009}^{+0.005}$ & $1.588_{-0.01}^{+0.009}$ & $1.59_{-0.02}^{+0.01}$ & $1.6_{-0.007}^{+0.011}$ \\
 &  & $\Gamma_{\rm FPMB}$ & $1.512_{-0.008}^{+0.006}$ & $1.589_{-0.015}^{+0.009}$ & $1.59\pm0.01$& $1.602_{-0.01}^{+0.011}$ \\
 &  & $kT$/keV & $400_{-300}^{+0}$ & $39\pm4$& $38_{-3}^{+4}$ & $50\pm5$\\
\hline\multirow{3}{1.5cm}{Disc} & \multirow{3}{2.1cm}{\textsc{relxilllpCp(1/2)}} & $R_{\rm in}/r_{\rm g}$ & $4.9_{-0.5}^{+2.3}$ & $5.4_{-0.7}^{+0.9}$ & $5.9_{-0.8}^{+1.2}$ & $5.3_{-0.5}^{+1.3}$ \\
 &  & $\theta/^\circ$ & $30_{-5}^{+4}$ & $34_{-1}^{+2}$ & $32_{-3}^{+2}$ & $33_{-1}^{+3}$ \\
&&$A_{\rm Fe}/A_{\rm Fe,\odot}$ & $4.0_{-0.7}^{+0.9}$ & $5.3_{-1.0}^{+2.0}$ & $7_{-2}^{+1}$ & $7_{-1}^{+2}$ \\
\hline\multirow{2}{1.5cm}{Lower reflection} & \multirow{2}{1.8cm}{\textsc{relxilllpCp(1)}} & $h/r_{\rm g}$ & $3.1_{-0.9}^{+1.9}$ & $4.6_{-0.7}^{+1.3}$ & $4.4_{-0.6}^{+0.9}$ & $3.63_{-0.05}^{+0.07}$ \\
 &  & $\log(\xi/{\rm erg\,cm\,s^{-1}})$ & $3.08_{-0.05}^{+0.14}$ & $1.9_{-0.5}^{+0.4}$ & $2.3\pm0.3$& $2.4_{-2.4}^{+0.1}$ \\
\multirow{2}{1.5cm}{Upper reflection} & \multirow{2}{1.8cm}{\textsc{relxilllpCp(2)}} & $h/r_{\rm g}$ & $100_{-40}^{+0}$ & $40_{-30}^{+60}$ & $40_{-20}^{+60}$ & $14_{-4}^{+9}$ \\
 &  & $\log(\xi/{\rm erg\,cm\,s^{-1}})$ & $0_{-0}^{+2}$ & $3.9\pm0.2$& $3.9\pm0.1$& $3.71_{-0.25}^{+0.08}$ \\
\hline\multicolumn{3}{c}{\multirow{2}{1.5cm}{$\chi^2/{\rm d.o.f.}$}}  & $676.7/624$ & $957.5/895$ & $890.1/864$ & $1267.9/1037$\\
 & &  & $1.08$ & $1.07$ & $1.03$ & $1.22$\\
\multicolumn{3}{c}{FPMA to FPMB comparison} & $1.09$ & $1.1$ & $1.0$ & $1.24$\\
\hline\\
  \hline
Component & Model & Parameter & \multicolumn{4}{c}{Epoch} \\\cmidrule(lr){4-7}& & & 5 & 6 & 7 & 8 \\\hline\multirow{4}{1.5cm}{Soft flux} & \multirow{2}{1.5cm}{\textsc{diskbb}} & Norm$_{\rm FPMA}$ & $2600_{-700}^{+2000}$ & $4000\pm2000$& $5000_{-2000}^{+4000}$ & $3000_{-1000}^{+2000}$ \\
 &  & $kT_{\rm FPMA}$/keV & $0.6_{-0.05}^{+0.03}$ & $0.56\pm0.05$& $0.53_{-0.05}^{+0.04}$ & $0.53_{-0.02}^{+-0.0}$ \\
 & \multirow{2}{1.5cm}{\textsc{diskbb}} & Norm$_{\rm FPMB}$ & $8000_{-4000}^{+13000}$ & $6000_{-4000}^{+16000}$ & $8000_{-4000}^{+14000}$ & $2900_{-1700}^{+500}$ \\
 &  & $kT_{\rm FPMB}$/keV & $0.48\pm0.05$& $0.48_{-0.07}^{+0.09}$ & $0.47_{-0.06}^{+0.05}$ & $0.5_{-0.05}^{+0.06}$ \\
\hline\multirow{5}{1.5cm}{Compton continuum} & \multirow{5}{2.1cm}{\textsc{relxilllpCp(1/2)}} & Norm$_{\rm FPMA}$ & $0.38_{-0.05}^{+0.11}$ & $0.4_{-0.1}^{+0.3}$ & $0.35347_{-0.10446}^{+7e-05}$ & $0.2535_{-0.0246}^{+0.0002}$ \\
 &  & $C_{\rm FPMB/FPMA}$ & $1.014\pm0.002$& $1.016\pm0.002$& $1.011\pm0.002$& $1.004_{-0.004}^{+0.003}$ \\
 &  & $\Gamma_{\rm FPMA}$ & $1.621_{-0.006}^{+0.008}$ & $1.65\pm0.01$& $1.658_{-0.012}^{+0.009}$ & $1.626_{-0.006}^{+0.008}$ \\
 &  & $\Gamma_{\rm FPMB}$ & $1.623_{-0.006}^{+0.008}$ & $1.65_{-0.01}^{+0.02}$ & $1.658_{-0.011}^{+0.009}$ & $1.63_{-0.002}^{+0.004}$ \\
 &  & $kT$/keV & $54_{-3}^{+9}$ & $80_{-20}^{+40}$ & $92_{-8}^{+76}$ & $230_{-80}^{+140}$ \\
\hline\multirow{3}{1.5cm}{Disc} & \multirow{3}{2.1cm}{\textsc{relxilllpCp(1/2)}} & $R_{\rm in}/r_{\rm g}$ & $5.8_{-0.4}^{+0.7}$ & $4.7_{-0.8}^{+1.5}$ & $4.4_{-0.7}^{+0.5}$ & $5.6_{-0.5}^{+2.6}$ \\
 &  & $\theta/^\circ$ & $31_{-2}^{+3}$ & $35_{-5}^{+4}$ & $38_{-4}^{+3}$ & $36_{-5}^{+4}$ \\
&&$A_{\rm Fe}/A_{\rm Fe,\odot}$ & $5.0_{-0.4}^{+3.7}$ & $8_{-3}^{+1}$ & $10_{-4}^{+0}$ & $6.0_{-1.0}^{+1.3}$ \\
\hline\multirow{2}{1.5cm}{Lower reflection} & \multirow{2}{1.8cm}{\textsc{relxilllpCp(1)}} & $h/r_{\rm g}$ & $3.6\pm0.4$& $3.1_{-0.5}^{+0.7}$ & $2.9\pm0.6$& $3.7_{-0.1}^{+2.0}$ \\
 &  & $\log(\xi/{\rm erg\,cm\,s^{-1}})$ & $1.7_{-0.7}^{+0.8}$ & $2.1_{-2.1}^{+0.4}$ & $2.1_{-2.1}^{+0.7}$ & $3.48_{-0.07}^{+0.14}$ \\
\multirow{2}{1.5cm}{Upper reflection} & \multirow{2}{1.8cm}{\textsc{relxilllpCp(2)}} & $h/r_{\rm g}$ & $16_{-5}^{+8}$ & $15_{-6}^{+12}$ & $6_{-2}^{+24}$ & $100_{-70}^{+0}$ \\
 &  & $\log(\xi/{\rm erg\,cm\,s^{-1}})$ & $3.49_{-0.04}^{+0.29}$ & $3.8_{-0.3}^{+0.1}$ & $3.9\pm0.1$& $0.3_{-0.3}^{+1.7}$ \\
\hline\multicolumn{3}{c}{\multirow{2}{1.5cm}{$\chi^2/{\rm d.o.f.}$}}  & $1136.9/995$ & $1033.7/909$ & $936.3/849$ & $1032.4/868$\\
 & &  & $1.14$ & $1.14$ & $1.1$ & $1.19$\\
\multicolumn{3}{c}{FPMA to FPMB comparison}  & $1.14$ & $1.12$ & $1.1$ & $1.13$\\
\hline\\
\end{tabular}
\end{table*}

We fit the data in \textsc{isis} \citep{houck00} version 1.6.2-41 across the full \nustar\ band, $3-78$\,keV, excluding $11-12$ and $23-28$\,keV due to sharp features which differ between FPMA and B, which we ascribe to instrumental effects \citep[these energies correspond to more variable regions of the empirical correction factor,][figure 5]{madsen15}. We give parameters in Table~\ref{tab:spec_fit}. Errorbars are given and plotted at the 90\% level for 1 parameter of interest. Residuals are shown in Figure~\ref{fig_res} and examples of the best-fitting models themselves are shown in Figure~\ref{fig_models}.
The evolution of the parameters is shown in Figure~\ref{fig_par_change}.

Due to slight calibration differences between FPMA and FPMB, we allow different \textsc{diskbb} parameters and photon indices ($\Gamma$)  between modules. We find that the typical difference in photon index is similar to the uncertainty in the fit, with FPMB always requiring a slightly harder model, though the difference is less than the stated calibration level \citep{madsen15}.
Similarly, FPMA always has a slightly hotter \textsc{diskbb} component.

\begin{figure}
\includegraphics[width=\columnwidth]{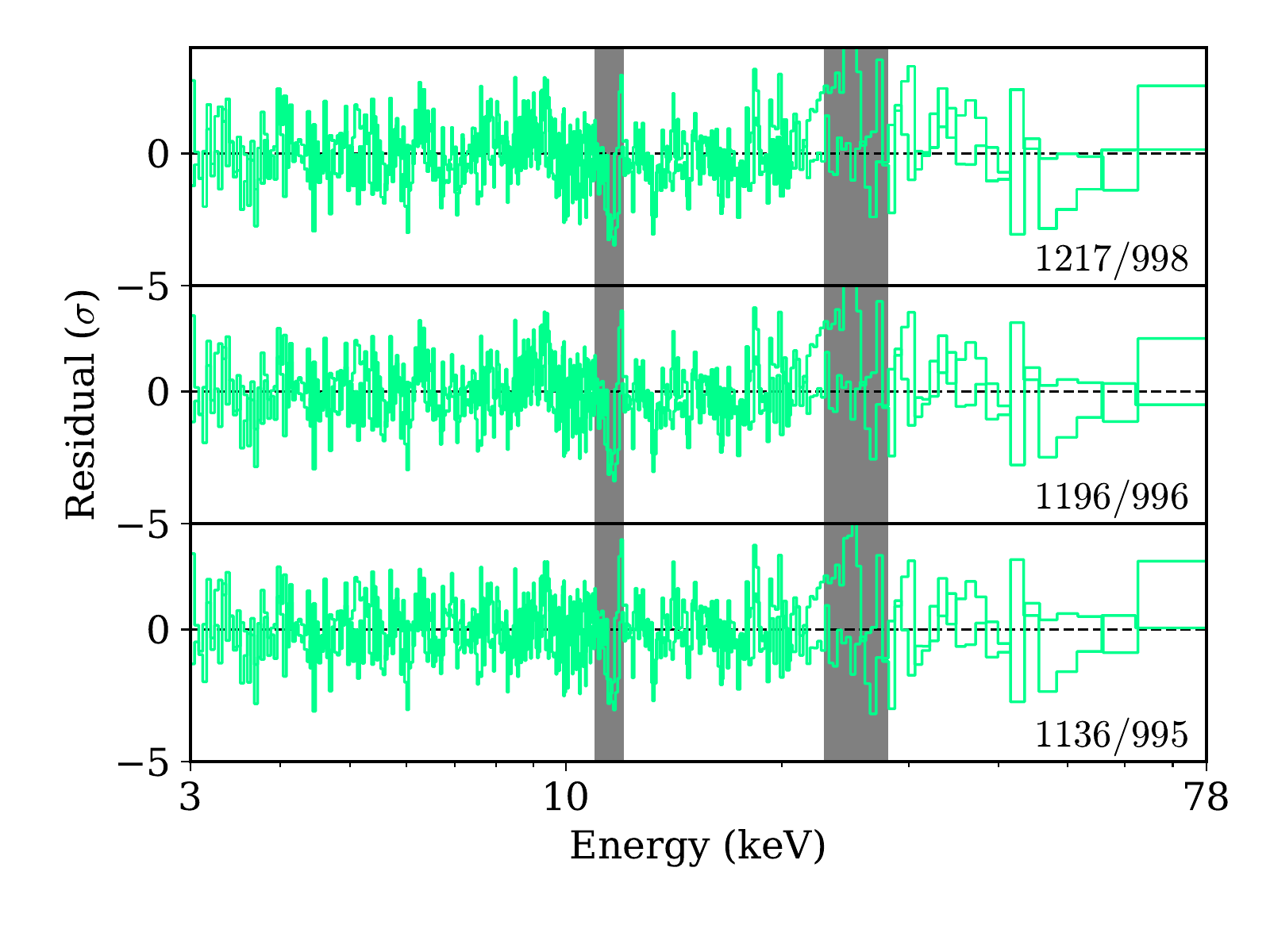}
\caption{Example of residuals for each variant of the model (see text for details). Top: single corona; middle: double corona with tied ionisation; bottom: final double corona model. The fit statistic/degrees of freedom is given in the lower right of each panel. All panels show epoch~5 only.}
\label{fig_model_res}
\end{figure}

To check that we require the two point extended corona, we also test a model with a single point source. This gives a significantly worse fit for each epoch: $\Delta\chi^2=17$ for 3 degrees of freedom in the weakest case, sometimes $\Delta\chi^2>100$. An example of the change in residuals for the different models is given in Figure~\ref{fig_model_res}.

For self-consistency, we tie black hole parameters and disc parameters that cannot change quickly between the two \textsc{relxilllpCp} components. We also use the self-consistent reflection fraction (the reflection strength is calculated based on the coronal height, \citealt{dauser16}), so include the continuum contribution from both components. However, as well as the height, we allow the disc ionisation to differ between the two components. We find that this difference is statistically necessary, with an average $\Delta\chi^2$ of 39 per epoch. This can be justified physically in several ways. The different heights in the different components mean that they mostly illuminate different regions of the disc (the lower component principally illuminates the inner disc). Alternatively, the variability in the system could be such that the flux from different coronal regions is dominant at different times: the disc ionisation could also change with this such that the ionisation when the upper corona is dominant differs from that when the lower dominates.
The resulting ionisation values often differ from the na\"{i}ve expectation that the lower corona should illuminate a more ionised inner disc. For this to be taken as physical, either some variability allows the disc to be less ionised when emission from the lower corona is dominant or a density gradient allows the more strongly illuminated inner region to have lower ionisation.
Alternatively, the relative ionisation values could be a modelling artefact; in this case, we can check the reliability of other parameters by tying both ionisation values together. Testing this on epoch 4 data (which has the strongest signal) retrieves parameters which are similar to (and in particular the inner radius is consistent at the 90\% level with) the values from the fits in Table~\ref{tab:spec_fit}.

\begin{figure}
\centering
\includegraphics[ trim={0.cm 0 0 0},width=\linewidth]{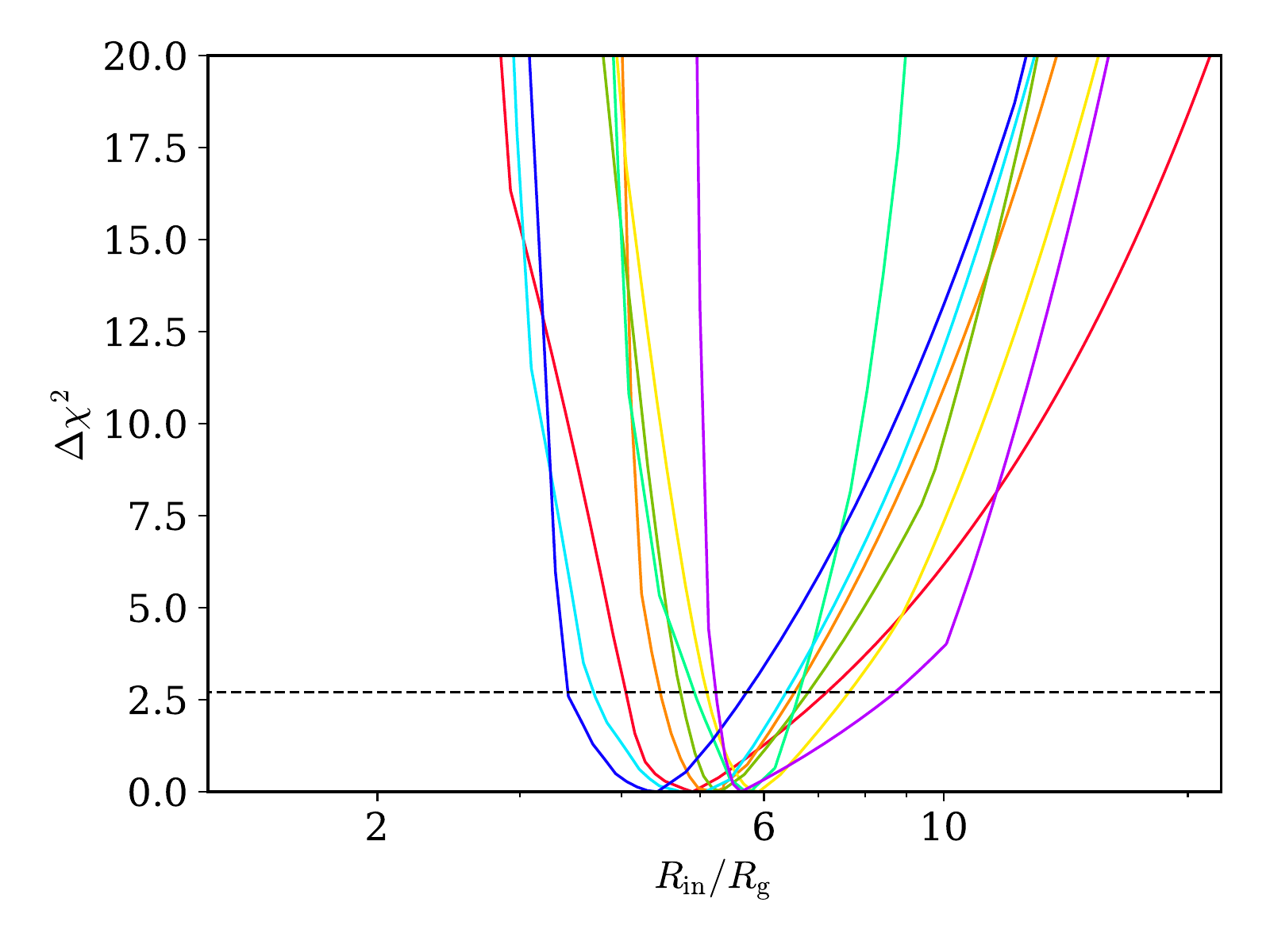}
\caption{Plot of constraint on inner disc radius, $R_{\rm in}$, in terms of change in fit statistic, $\Delta\chi^2$, for each epoch. The dashed line indicates the change in $\chi^2$ corresponding to a 90\% confidence interval.
}
\label{fig_rin_chi2}
\end{figure}

Owing to the strong degeneracy between black hole spin and disc truncation, we fit for inner radius, $R_{\rm in}$, in a maximally spinning (dimensionless spin parameter $a=0.998$) space-time. The resulting values show little spread around their weighted mean, $R_{\rm in} \sim 5.3\,r_{\rm g}$ ($r_{\rm g}=GM_{\rm BH}/c^2$), see Figure~\ref{fig_rin_chi2}. If this radius is $R_{\rm ISCO}$, it implies a low spin black hole.
A full estimate of the black hole spin, including low-energy data from \nicer, will be presented in a forthcoming paper (Fabian et al., in prep.).

The iron abundance of the disc is found to be significantly higher than solar ($A_{\rm Fe}\sim5A_{\rm Fe,\odot}$). This is not necessarily unexpected, since stars vary in metallicity, but the value found is likely to be an overestimate, particularly given the ubiquity of apparent super-solar iron abundances \citep{garcia18}. The over-estimate could be due to a higher density disc (as predicted for stellar mass black holes, \citealt{svensson94,garcia16}) which would show stronger iron lines at a given metallicity \citep{garcia16,tomsick18,jiang19}. The difference in density should not have a strong effect on other parameters of the system (Jiang et al. MNRAS accepted). Additionally, high metallicity could occur if the supernova which formed the black hole polluted the surface of the companion with metal-rich material, which is now being accreted.

Many of these fits are formally poor, in the sense of having low null hypothesis probabilities. However, the statistical errors in the spectrum are comparable to the calibration precision of \nustar\ due to the extremely high signal in the datasets used here, so calibration differences between the detectors may lead to inflated $\chi^2$ values. To give a guide to how significant this effect is, we also show the value:\textbf{
$$\frac{\left((D_{\rm A}-M_{A})-(D_{\rm B}-M_{B})\right)^2}{\left(E_{\rm A}^2+E_{\rm B}^2\right)\times{\rm d.o.f.}}$$}
where $D_{i}, M_{i}, E_{i}$ are the data, model and error values respectively for detector $i$ and ${\rm d.o.f.}$ is the number of degrees of freedom, i.e. the number of bins minus the 6 variables in our model which can differ between detectors (4 from two instances of \textsc{diskbb}; the normalisation difference; and the difference in $\Gamma$). This is essentially a reduced $\chi^2$ value testing that FPMA matches FPMB. All values are similar to the reduced $\chi^2$ found for the respective source model. Since the model cannot simultaneously match both detectors better that the detectors match each other,  this justifies the fit quality of the source models.

Another way of determining the effects of calibration uncertainties is to add a systematic error to the measurement uncertainties; here, a systematic error of below 0.5\% brings the reduced $\chi^2$ to unity; the effects on parameters of interest are minor. Since the effects of systematic error are binning dependent and the level chosen is somewhat arbitrary, we consider the parameters derived without addtion of systematic error for the rest of this work.

\begin{figure}
\centering
\includegraphics[width=\columnwidth]{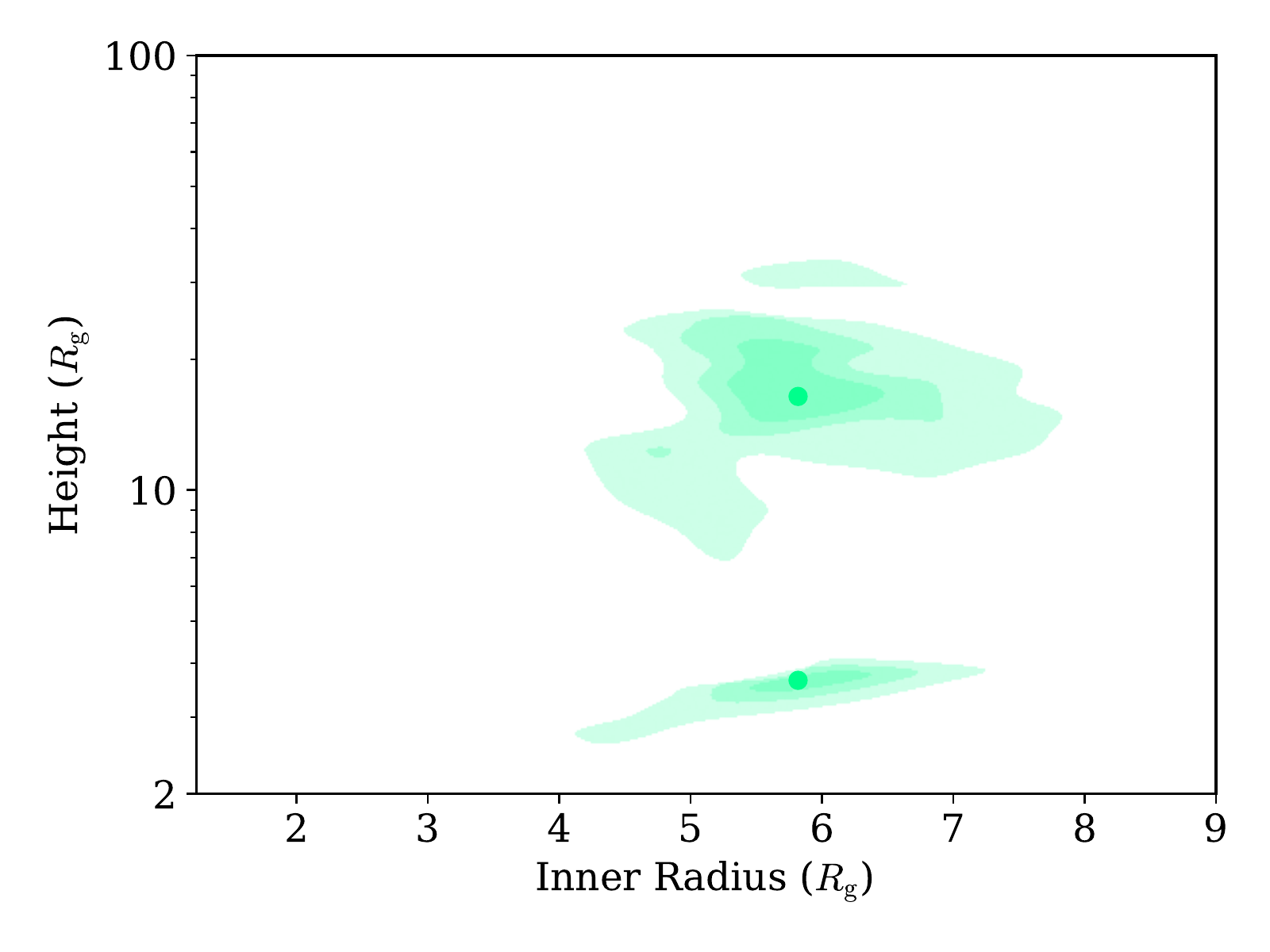}
\caption{Contour plots for the change in fit statistic with coronal height (for each of the two components) and disc inner radius. The colour levels indicate $1\sigma$, 90\% and $3\sigma$ confidence. For clarity, only epoch~5 is shown as an example.}
\label{fig_cont}
\end{figure}

We also check for the influence of strong degeneracies on the measured parameters, in particular the height of each coronal component and the disc inner radius. We show the confidence contour plot of $R_{\rm in}$ against the height of each coronal component in Figure~\ref{fig_cont}. This shows that while there is a mild degeneracy between inner radius and lower coronal height, each parameter is independently well constrained.

Various parameters (e.g. coronal temperature) change significantly between epochs. How these changes are related, to each other and to properties of the rapid variability, is considered further in Section~\ref{sec_comparison}.

\subsection{Variability analysis: power spectra}
\label{sec:psds}

\begin{figure}
\centering
\includegraphics[ trim={0.cm 0 0 0},width=\linewidth]{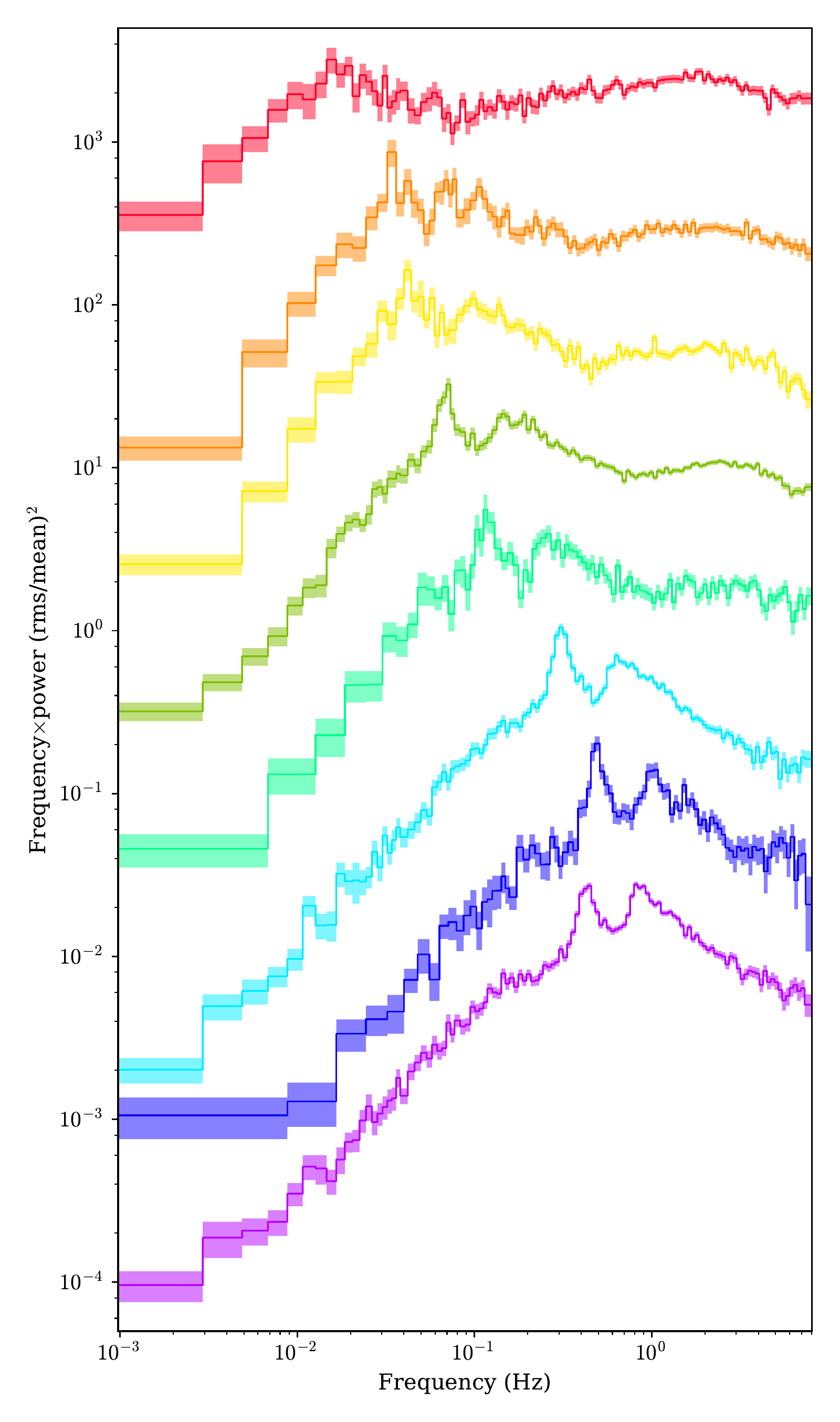}
\caption{PSDs of \nustar\ data, with the RMS normalisation. Successive PSDs are offset by a factor of 5. Poisson noise has been subtracted based on the best-fitting values and each PSD has been rebinned to a geometric progression of at least 1.05 for clarity. Frequencies of features in the PSD (QPO and low-frequency break) increase over the first section of the outburst. During the latter stages of the outburst, the variability decreases.
}
\label{fig_nustar_psds_allobs}
\end{figure}

The timescales on which a source varies may be quantified with the Power Spectral Distribution (PSD, e.g. \citealt{priestley81})

$$P(f) = |A(f)|^2$$

where $A(f)$ is the Fourier transform of the flux at frequency $f$.

Initially, we produce periodograms from the full calibrated \nustar\ band ($3-78$\,keV), using lightcurve segments of 1024\,s with $0.0625$\,s $=1/16$\,s bins.
We then produce PSDs from the average of all periodograms in an epoch, binning frequencies if necessary to ensure that each PSD data point is produced from at least 20 periodogram values (so that the error on the PSD value is approximately Gaussian). We estimate the size of the error of each PSD point from the variance of the periodogram values which produce it.

These PSDs are shown for each epoch in Figure~\ref{fig_nustar_psds_allobs}. The low-frequency cut-off in power increases in break frequency as the outburst progresses. Additionally, a QPO is present close to the break frequency in each observation (although its detection is very marginal in the first); a further peak is present close to double the primary QPO frequency. These QPO frequencies also (with the exception of the final observation) increase with time.

We also test for changes in variability properties with energy by splitting each light curve into 5 energy bands (3-5, 5-6, 6-9, 9-13 and 13-78\,keV) with approximately equal counts. This shows a similar PSD shape in each band and only a slight change in variability amplitude. Therefore, we consider only the full band PSDs here (a detailed analysis of the changes with energy will be presented as part of a future work).

\subsubsection{Fitting}

Power spectra of accreting black holes can typically be fit with the sum of several Lorentzians \citep{olive98,belloni02}. We fit such a model, typically using 5 Lorentzians (apart from epochs 1 where only 3 are necessary, 4 where 6 are necessary and 7 where 4 are necessary), and including an additional constant (independent of frequency) component for the Poisson noise. We fit the two FPMs as separate datasets with the same source model but independent Poisson noise components.

We then use Markov-Chain Monte-Carlo methods to fit each of the PSDs, utilising the \textsc{XSPEC\_EMCEE} implementation\footnote{Written by Jeremy Sanders, based on the \textsc{EMCEE} package \citep{foreman13}.}. We use 150 walkers for 5000 steps after a burn in period of 1000 steps. 
For each parameter, we apply a simple uniform prior across a range determined by eye to encompass all reasonable values.

\begin{table}
\caption{Parameters of QPO fundamental for each epoch analysed. Epoch~1 does not have a clear detection of the QPO, so is not included.}
\label{tab_qpo_pars}
\centering\begin{tabular}{cll}
\hline
Epoch & $\nu_{\rm peak}$ & $Q$ \\
\hline
2 & $0.036_{-0.002}^{+0.002}$ & $3.1_{-0.8}^{+1.3}$ \\
3 & $0.041_{-0.002}^{+0.003}$ & $2.5_{-0.5}^{+1.6}$ \\
4 & $0.068_{-0.001}^{+0.001}$ & $7.7_{-1.5}^{+2.2}$ \\
5 & $0.117_{-0.006}^{+0.006}$ & $6.5_{-3.0}^{+7.6}$ \\
6 & $0.308_{-0.002}^{+0.002}$ & $8.5_{-0.7}^{+0.8}$ \\
7 & $0.489_{-0.006}^{+0.007}$ & $10.5_{-2.3}^{+2.9}$ \\
8 & $0.428_{-0.004}^{+0.036}$ & $7.9_{-4.9}^{+1.1}$ \\
\hline
\end{tabular}
\end{table}

We then extract characteristic time scales to compare with spectral properties. We extract the QPO peak frequency ($\nu_{\rm peak}=\sqrt{\nu_0^2+\sigma^2}$) by finding the narrow Lorentzian with the highest normalisation close to the visible QPO peak frequency (where the allowed range is determined by eye to exclude adjacent peaks). We define a Lorentzian as narrow based on the standard threshold, $Q>2$, where $Q=\sigma/2\nu_0$ is the quality factor. Epoch~1 does not have a clear QPO, so we do not extract a QPO frequency for it. We show the derived parameter values in Table~\ref{tab_qpo_pars}.
We note that the low-frequency break follows the change in QPO frequency; due to the similar changes in each characteristic frequency and the complication in reliably extracting the low-frequency break from the multiple Lorentzian model, we do not consider the low-frequency break quantitatively.

\section{Comparison of parameter evolution and discussion}
\label{sec_comparison}

\begin{figure*}
\centering
\includegraphics[ trim={0.cm 0 0 0},width=\linewidth]{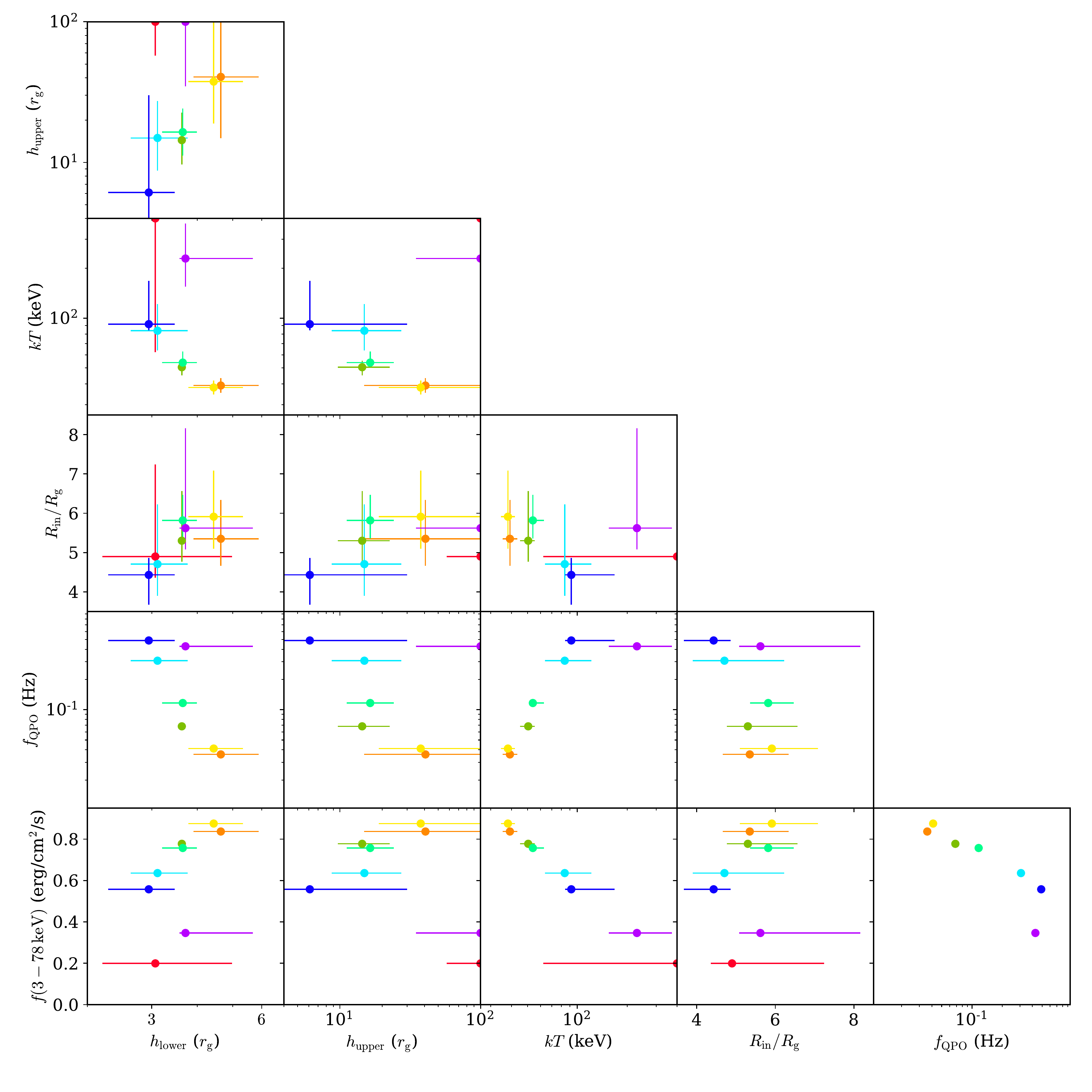}
\caption{Plot comparing evolution of different parameters. the colour of each epoch matches previous figures. Correlations are present between various parameters -- see text for details.}
\label{fig_par_cors}
\end{figure*}

\begin{figure}
\centering
\includegraphics[ trim={0.cm 0 0 0},width=\linewidth]{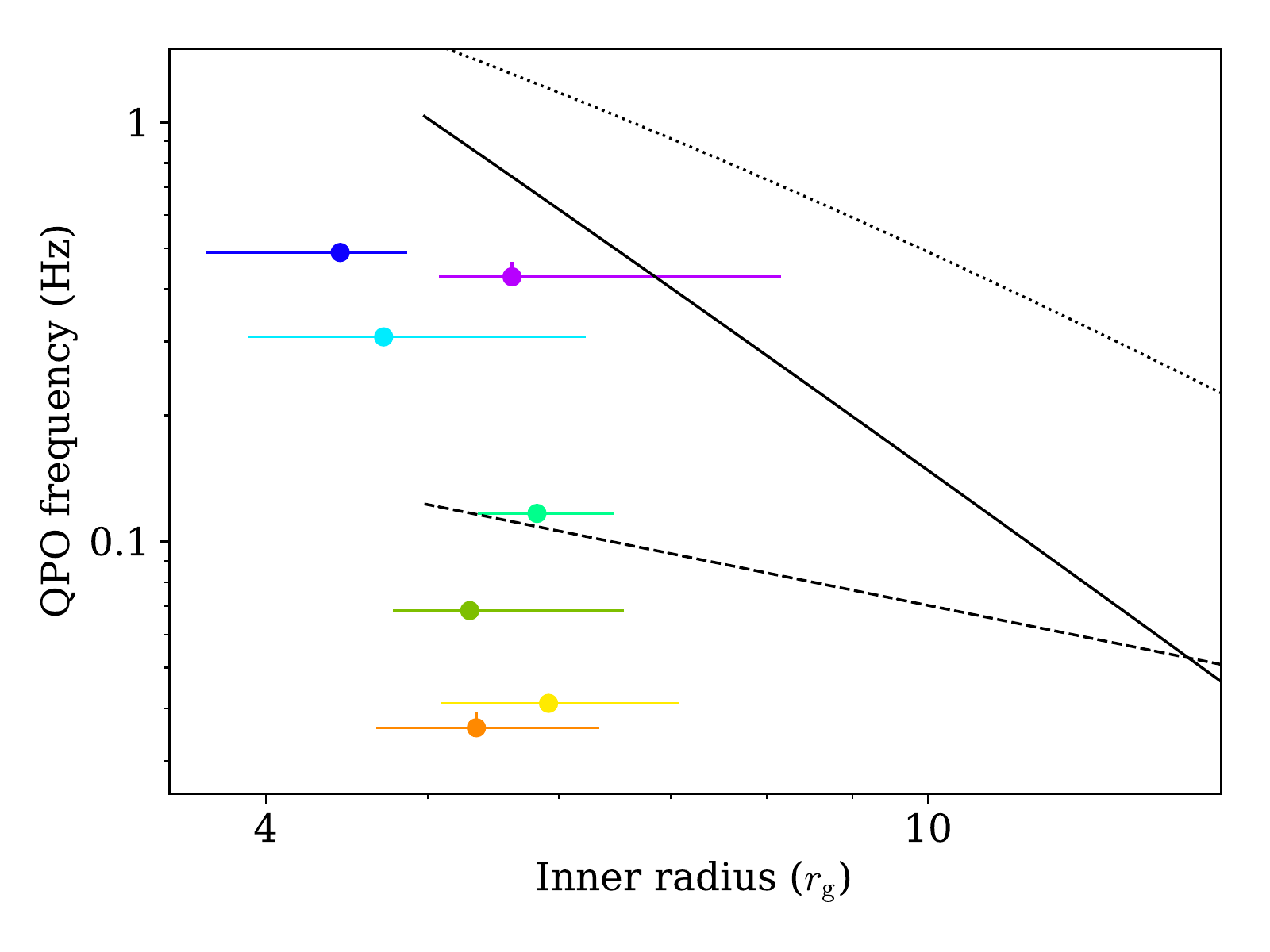}
\caption{Comparison of relation between inner radius and QPO frequency with various models. Solid: Lense-Thirring frequency of particle at $R_{\rm in}$. Dotted: solid-body precession  of hot flow extending from $R_{\rm ISCO}$ to $R_{\rm in}$. Dashed: Global Normal Disk Oscillation (see text for details of each model). To reproduce the observed range of QPO frequencies, all these models require a significantly greater change in inner disc radius than is measured.}
\label{fig_qpo_rin}
\end{figure}

\begin{figure*}
\centering
\includegraphics[ trim={0.cm 0 0 0},width=0.666\columnwidth]{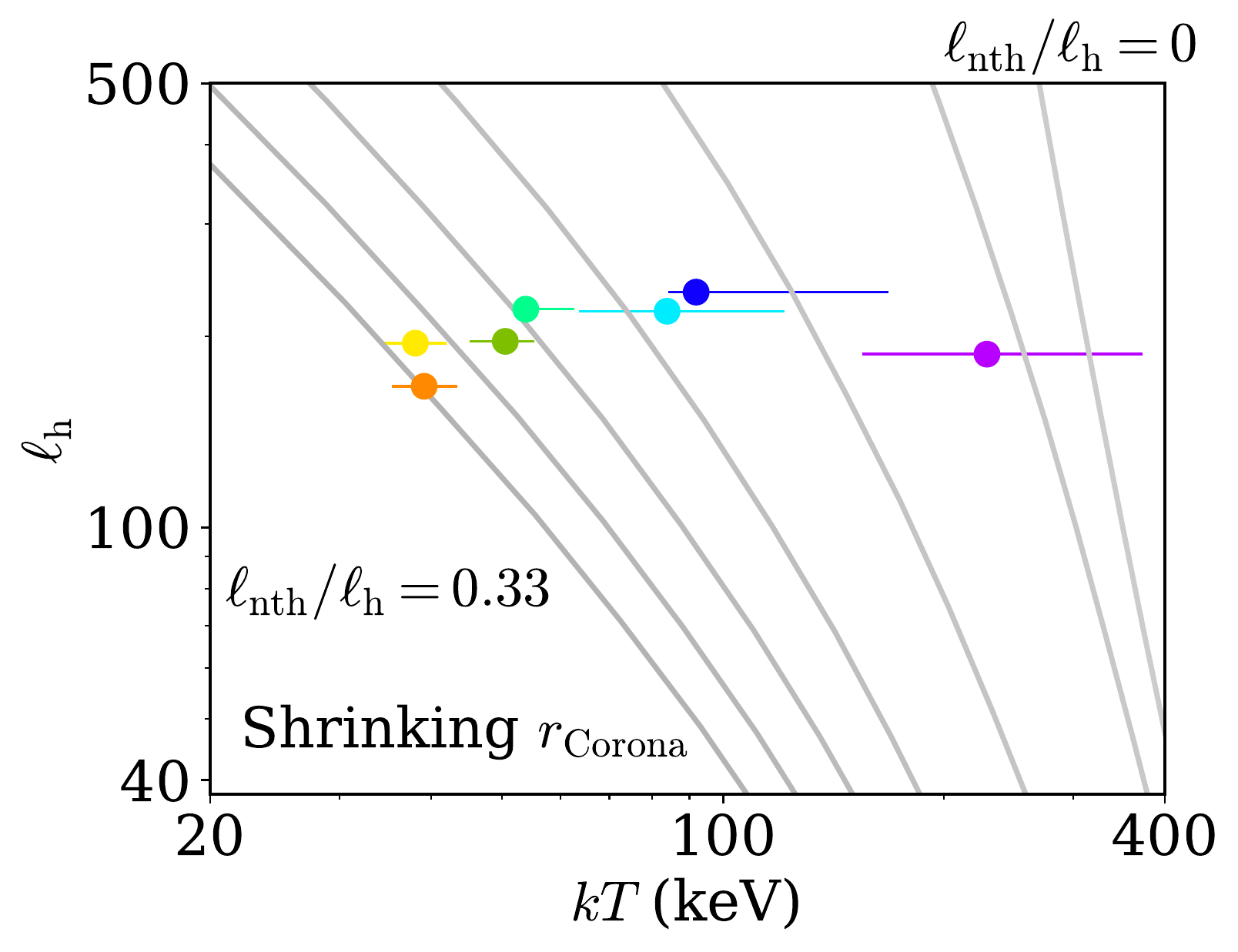}
\includegraphics[ trim={0.cm 0 0 0},width=0.666\columnwidth]{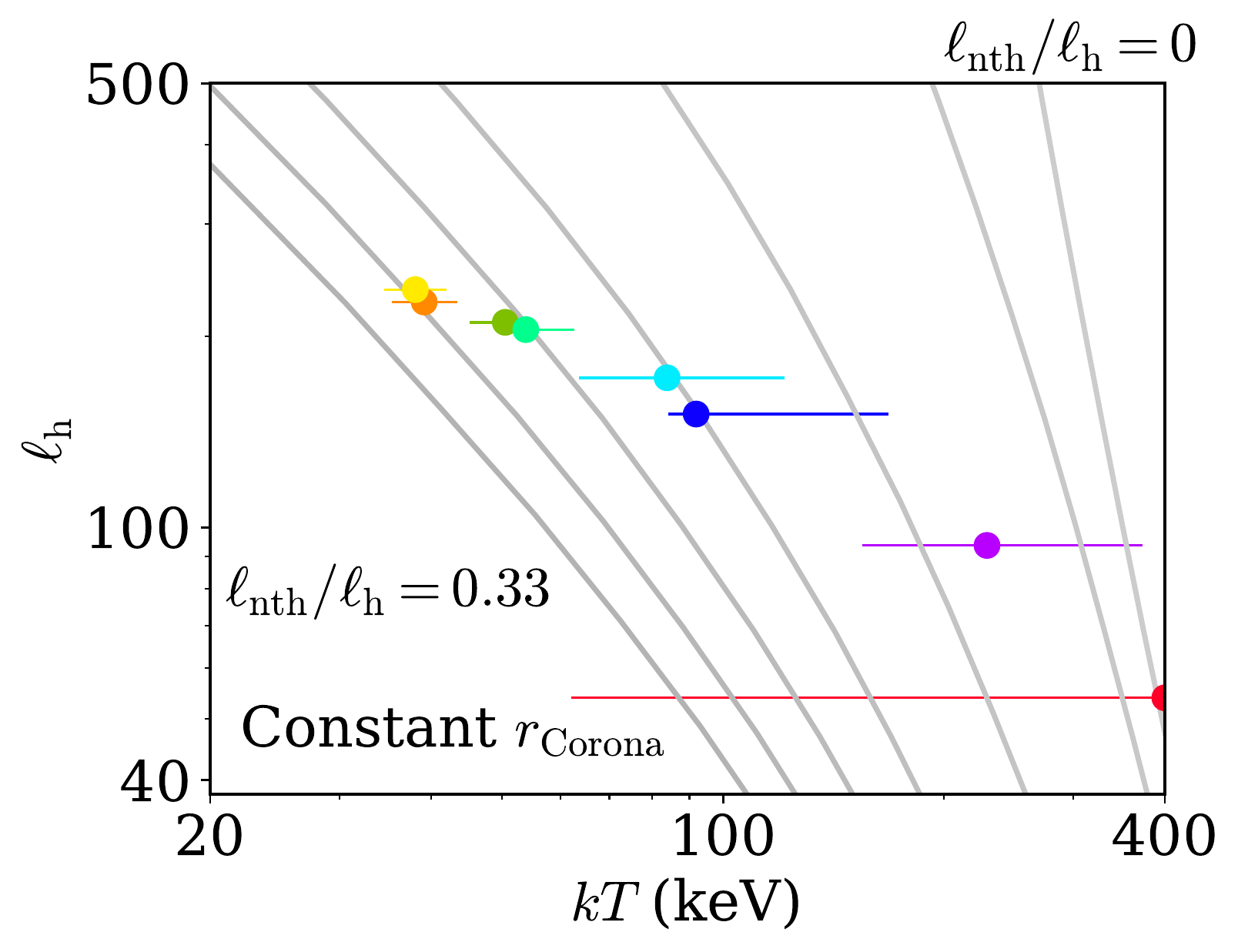}
\includegraphics[ trim={0.cm 0 0 0},width=0.666\columnwidth]{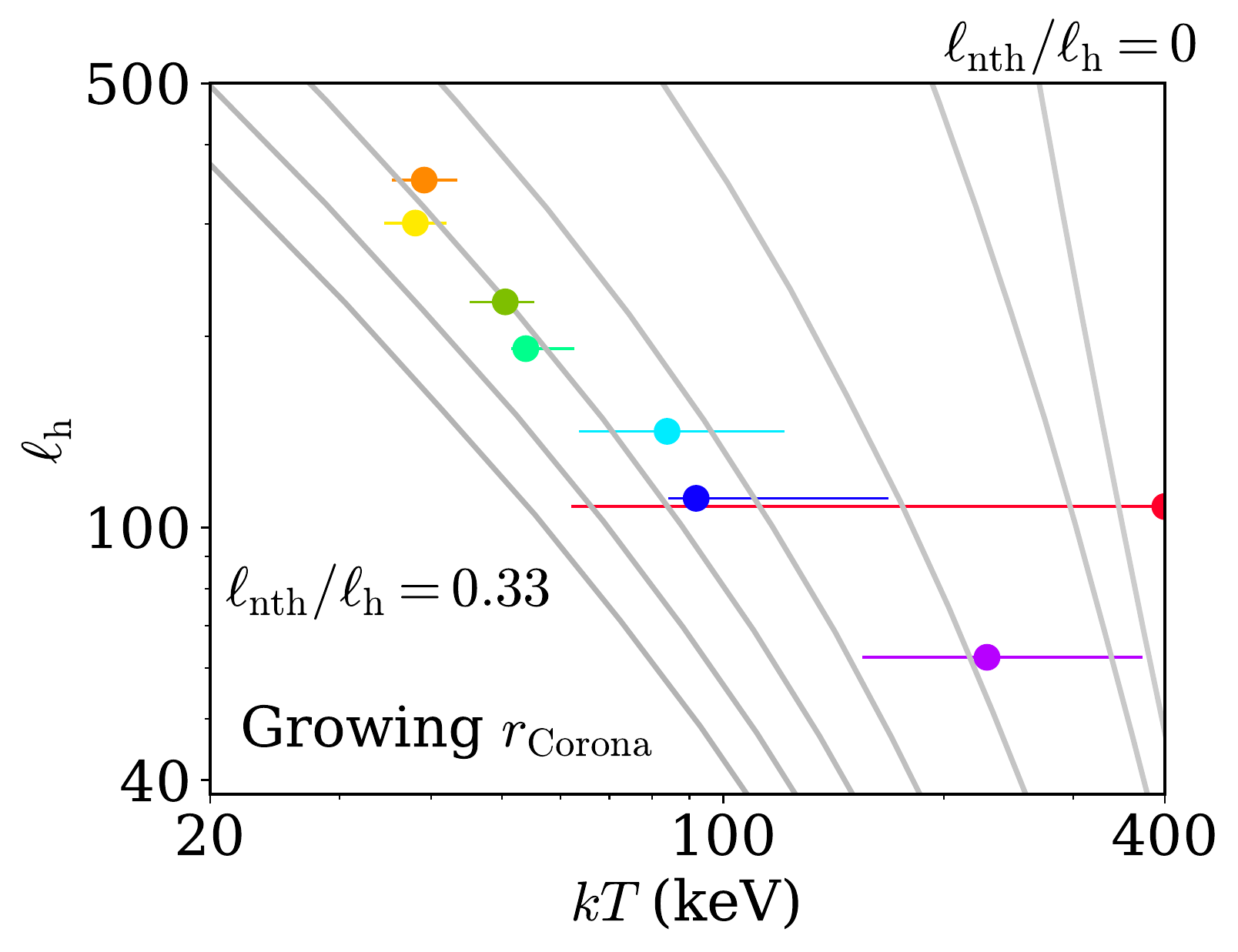}
\caption{Coronal compactness compared with coronal temperature at each epoch. Theoretical  curves of constant non-thermal fraction(grey lines) are taken from \citet{fabian17}; these have, from right to left, $\ell_{\rm nth}/\ell_{\rm h}=0,0.01,0.09,0.17,0.23,0.29,0.33$. Data from each epoch have the same colours as other figures. Errors in $\ell_{\rm h}$ are dominated by the choice of coronal radius so error bars are not shown; instead, different choices are given in the different panels. The left hand panel uses a coronal radius decreasing linearly from $10r_{\rm g}$ to $5r_{\rm g}$; the centre panel uses a constant coronal radius of $10\,r_{\rm g}$; and the right hand panel uses a coronal radius which increases from $5\,r_{\rm g}$ to $15\,r_{\rm g}$ (see text).}
\label{fig_lt}
\end{figure*}

We have described the evolution of the hard X-ray emission from the initial hard state of \maxi. This is a powerful probe of the inner accretion system and our analysis shows a reflecting inner disc extending close to the ISCO of the central black hole throughout the hard state of the outburst. This is in stark contrast to the large change in variability timescale, which increases by a factor of around $30$.
We also find changes in the coronal temperature and extent.

We compare how key parameters evolve with respect to each other in Figure~\ref{fig_par_cors}.
While many pairs of parameters appear uncorrelated, there are correlations between the coronal temperature and each of QPO frequency and flux.
We discuss possible resaons for the changes (or lack thereof) of system properties and their correlations in the following subsections.

\subsection{Disc inner radius}

One important issue which is not yet fully resolved is when during an outburst the disc is truncated.
Here, we have shown several observations in the hard state without strong disc truncation.
To be consistent with observations of truncated discs at low accretion rate \citep{tomsick09} this implies that the disc fills (the inner radius moves inwards) during the initial rise: most of the observations used here are at or after the peak of the outburst.
This would imply that discs can reach close to the ISCO regardless of hard/soft state, in this case showing bright hard state emission with a reflecting disc at or close to the ISCO.

\subsection{QPO frequency}

QPOs are often found in the power spectra of X-ray binaries, though their origin is not yet fully understood.
Such QPOs are a rapidly oscillating change in the flux of the source. They are observed to occur principally in the coronal powerlaw emission \citep{rodriguez02,casella04} but most explanations invoke some link to the disc, as the disc possesses more accessible characteristic timescales, particularly those associated with the inner edge.

Various models associate the inner disc radius with the characteristic scale which produces QPOs.
We plot our measurements of QPO frequency and inner radius along with some models in Figure~\ref{fig_qpo_rin} \citep[similarly to][]{fuerst16}.
Firstly, relativistic effects introduce various precession frequencies. Of these, Lense-Thirring (nodal) precession is most likely to lie in the frequency range of LFQPOs \citep{stella98,stella99}. Following \citet{ingram09}, we plot the frequencies of a single particle and a hot flow extending to $R_{\rm ISCO}$ (we do not show a hot flow with the inner radius set by bending waves as this radius is always larger than our measurements). For illustration, we take a black hole mass $M_{\rm BH}=100\,{\rm{M_{\odot}}}$ (a high mass is required to have low enough frequencies); a radial surface density profile ($\Sigma\propto r^{-\zeta}$) having $\zeta=0$ to match simulations \citep{fragile07}; and choose $a=0.3$ to give similar frequencies to those observed while not having the measured $R_{\rm in}<R_{\rm ISCO}$.
Another possibility is an oscillation mode of the disc, such as the global normal disc oscillation discussed in \citet{titarchuk00}: an oscillation of the whole disc in the direction normal to the disc plane. We also plot this in Figure~\ref{fig_qpo_rin}, again taking $M_{\rm BH}=100\,{\rm{M_{\odot}}}$ and choosing the outer disc radius, $r_{\rm out}=10^4r_{\rm g}$ to give reasonable frequencies.
All of these models require a much greater change in inner radius than is measured to explain the range of QPO frequencies.
Therefore, either some other process governs the frequency of QPOs or the inner radius of the reflecting material does not match the edge of the oscillating material.

Our results show a change in QPO frequency without a significant change in disc inner radius \citep[see also][]{fuerst16,xu17}, which is a challenge for models which rely on geometric (orbital or precession) timescales related to the inner edge of the disc.
Since QPOs appear to have different observed properties depending on inclination \citep{vandeneijnden17}, some geometric effects are likely; these could still occur but be linked to the frequency differently, such as  jet precession, or indirectly, such as a coronal oscillation which is directed parallel to the plane of the disc.

There are other possible models for QPO production: it has also been suggested that feedback between coronal heating of the disc and increased seed photon rates could have resonant frequencies which manifest as QPOs.

Alternatively, the QPO could be generated directly by oscillations in the corona \citep[e.g.][]{cabanac10,zanotti05}, such as a resonant mode of the constituent plasma.
A simple prescription to describe this could be a sound wave passing across the corona. The frequency then scales as
$$ \nu = A \frac{c_{\rm s}}{2d} \simeq A \frac{T_{9}^{1/2}}{d_2}\ {\rm Hz}$$
where $c_{\rm s}$ is the sound speed, $T=10^{9}T_{9}$\,K is the temperature, $d=100d_{2}r_{\rm g}$ is the distance across the corona and $A$ is a factor of order unity.
This would fit with the change in coronal extent and temperature implied by the spectral fitting -- in the smaller, hotter corona, oscillations would have a higher frequency. More quantitatively, during the outburst the coronal height reduces by around a factor of 10 and the temperature increases by a factor of at least 4. This would increase the associated frequency by a factor of $\sim20$, similar to the observed increase in QPO frequency.
The average value of $A$ for $\nu$ to match $\nu_{\rm QPO}$ is then $\sim1/30$ (taking $M_{\rm BH}=10\,{\rm M}_\odot$).
This factor could include contributions from the turnaround time at each end of the corona or from other physical processes. MHD calculations \citep{edwin83} show that magnetic fields affect the frequency of various modes of oscillation.
Detailed calculations of expected values of $A$ are beyond the scope of this work.

\citet[][figure~10]{remillard06} show that QPO frequency correlates with disc flux in hard/steep powerlaw intermediate states. The observations presented here have a weaker disc component which is not unambiguously detected but may be clearer at soft energies, so this could be investigated for example with \nicer.

While the QPO is most dominant at high energies, it is also strongly detected in the 3-10\,keV band, which is also covered by \nicer. Owing to the more frequent coverage of \maxi\ by \nicer, a more detailed analysis of the QPO progression can be made with these data.

\subsection{Coronal temperature}

The corona also shows changes in its mean properties: the temperature anti-correlates with flux. This has been observed in other individual XRBs \citep{joinet08,motta09} and AGN \citep{lubinski10}. This could happen because pair production from photon collisions is regulating the temperature \citep{svensson84,zdziarskii85,fabian15}: at higher fluxes, there are more photons, which allows sufficient pairs to be produced at lower temperatures.
The coronal temperatures observed here are allowed by the pair thermostat, being below the pair-production limit which is observed to limit accreting sources as a population \citep{fabian15}.
The lower temperatures than the pair limit can be explained by a deviation from a pure thermal distribution, which is expected as the cooling timescale is less than the collision timescale \citep{fabian17}.
The presence of a non-thermal tail to the particle distribution reduces the critical temperature \citep{fabian17}.
These considerations do not take sub-structure in the corona into account; this remains a potential caveat when estimating the coronal compactness from its total luminosity and size.

We consider this correlation in more detail by comparing the corona's radiative compactness with its electron temperature.
We take electron temperatures from our fits and calculate the coronal compactness following the methodology of \citet{fabian15}. We take the luminosity from the total direct flux of our best-fit models at a distance of 3.5\,kpc. Calculation of the compactness also requires a radius; the coronal prescription used here (including contributions from two points) does not readily convert to an equivalent spherical size so we try several prescriptions. Firstly, we consider a fiducial $10r_{\rm g}$ size for all observations. We also consider the effect of a shrinking radius, as could be implied by the reduction in illumination of the outer disc with time. We reduce the radius linearly by epoch from $10r_{\rm g}$ to $5r_{\rm g}$, guided by the fractional change in the lag amplitudes from \citet{kara19} (since the lags are driven by the location of the majority of flux).
We plot these measurements in Figure~\ref{fig_lt}. The constant coronal size prescription gives a smooth trend but does not align with an obvious physical locus (such as constant non-thermal fraction, $\ell_{\rm nth}/\ell_{\rm h}$). The shrinking corona has an approximately constant compactness for all epochs after the peak (i.e. not the first epoch). With these size prescriptions, the non-thermal fraction is higher at high flux.
We could instead assume consistent physical conditions within the corona, manifesting as a constant non-thermal fraction, and use this to infer a trend in effective coronal radius: increasing $r_{\rm Corona}$ from 5 to $15\,r_{\rm g}$  gives a roughly constant non-thermal fraction of around 20\%.
This would imply an anti-correlation between the vertical coronal extent and the effective coronal radius, so that the corona had changed shape from prolate or cylindrical to oblate.
If this is the case and the QPO frequency is associated with coronal size, then the observed trend in QPO frequency would imply that the vertical, rather than horizontal, extent is the relevant dimension.
We stress that these relations depend strongly on the assumed prescription for any change in coronal radius, so must be treated with caution.

\label{section_discussion}

This paper covers only a small part of the data available on this outburst: further work on this and similar outbursts with the new generation of facilities now available will surely help to resolve these outstanding questions.

\section{Conclusions}
\label{section_conclusions}

We have described the evolution of X-ray spectral and timing properties of the recent outburst of \maxi\ during the hard state. In particular:
\begin{itemize}
\item spectral features change subtly: the broad component of the iron line remains almost constant while the narrow core reduces with time;
\item reflection modelling implies a small inner radius in all observations, consistent with the ISCO of a low to moderate spin black hole ($\sim5.3r_{\rm g}$);
\item the change in the iron line shape can be explained by coronal emission being more concentrated close to the black hole in later observations;
\item the coronal temperature is higher at lower flux. It can be explained by pair-production if the effective radius of the corona grows with time and the non-thermal electron fraction is $\sim20\%$;
\item characteristic frequencies of the variability, QPO frequency and low-frequency cut-off, increase by a factor of $\sim30$ during this outburst state;
\item and the change in QPO frequency with stable $R_{\rm in}$ is hard to reconcile with hot inner disc precession models for QPOs. It is probably related to the temporal development of the corona.
\end{itemize}

\section*{Acknowledgements}

We thank Fiona Harrison for approval of these DDT observations and Karl Forster for their prompt scheduling.
DJKB acknowledges financial support from the Science and Technology Facilities Council (STFC).
ACF acknowledges support from the ERC Advanced Grant FEEDBACK 340442.
AWS is supported by an NSERC Discovery Grant and a Discovery Accelerator Supplement.
This work made use of data from the \nustar\ mission, a project led by the California Institute of Technology, managed by the Jet Propulsion Laboratory, and funded by the National Aeronautics and Space Administration. This research has made use of the \nustar\ Data Analysis Software (NuSTARDAS) jointly developed by the ASI Science Data Center (ASDC, Italy) and the California Institute of Technology (USA).

\bibliographystyle{mnras}
\bibliography{maxij1820}

\bsp	
\label{lastpage}
\end{document}